%% file: CSSB_HMM_SysRisk10.tex
\documentclass[12pt]{article}
\input{preamble}

\begin{document}
\title{\bfseries ``Speculative Influence Network'' during financial bubbles: application to Chinese Stock Markets}
\author{L. Lin{\small$^{\mbox{\,\ref{ECUST}, \ref{ECUSTFin}~\footnote{Email: llin@ecust.edu.cn}}}$} 
and 
D. Sornette{\small$^{\mbox{\,\ref{ETH}, \ref{SFI}~\footnote{Email: dsornette@ethz.ch}}}$} 
}%
\maketitle

\vspace{-9mm}

\begin{enumerate}
\item{School of Business, East China University of Technology and Science, 200237 Shanghai, China}\vspace{-3mm}\label{ECUST}%
\item{ETH Zurich, Department of Management, Technology and Economics, Scheuchzerstrasse 7, CH-8092 Zurich, Switzerland}\label{ETH} \vspace{-3mm}%
\item{Swiss Finance Institute, c/o University of Geneva, 40 blvd. Du Pont d'Arve, CH 1211 Geneva 4, Switzerland}\label{SFI}\vspace{-3mm}%
\item{Research Institute of Financial Engineering, East China University of Technology and Science, 200237 Shanghai, China}\label{ECUSTFin}
\end{enumerate}

\abstract{We introduce the Speculative Influence Network (SIN) to decipher the causal relationships 
between sectors (and/or firms) during financial bubbles. The SIN
is constructed in two steps. First, we develop a Hidden Markov Model (HMM) of 
regime-switching between a normal market phase represented by a geometric Brownian motion (GBM)
and a bubble regime represented by the stochastic super-exponential Sornette-Andersen (2002) bubble model.
The calibration of the HMM provides the probability at each time for a given security to be in the bubble regime.
Conditional on two assets being qualified in the bubble regime, we then use 
the transfer entropy  to quantify the influence of the returns of one asset $i$ onto another asset $j$,
from which we introduce the adjacency matrix of the SIN among securities.
We apply our technology to the Chinese stock market during the period 2005-2008, during which a normal phase
was followed by a spectacular bubble ending in a massive correction. We introduce 
the Net Speculative Influence Intensity (NSII) variable as the difference between the transfer entropies 
from $i$ to $j$ and from $j$ to $i$, which is used in a series of rank ordered regressions to predict the 
maximum loss (\%{MaxLoss}) endured during the crash. The sectors that 
influenced other sectors the most are found to have the largest losses. There is a clear prediction skill obtained
by using the transfer entropy involving industrial sectors to explain the 
\%{MaxLoss} of financial institutions but not vice versa.  We also show that the bubble state variable
calibrated on the Chinese market data corresponds well to the regimes when the market exhibits a strong
price acceleration followed by clear change of price regimes.
Our results suggest that SIN may contribute significant skill to
the development of general linkage-based systemic risks measures and early warning metrics.
 }

\section{Introduction}

It is widely recognized that the backdrop of almost every proceeding financial bubble is the prevalence of speculative mania, which causes valuation to drift out of whack, associated with a reinforcing imbalance between the size of unrealized supply and 
demand intentions, which forms the genesis of the potential market collapse. 
Speculative mania is by and large embodied in a variety of herding behaviors when investors imitate and follow other investors' strategies while tending to suppress their own private information and beliefs \citep{Devenow1996,Avery1998}. Such imitation can be either rational or irrational. Rational herding results from different possible mechanisms, such as (i) the anticipation by rational investors concerning noise trader's feedback strategies  \citep{Delong1990}, (ii) Ponzi schemes resulting from agency costs,
(iii) monetary incentives given to competing fund managers \citep{Dimi2004,Dass2008} and (iv) rational imitation in the presence of uncertainty  \citep{Didier2000}. In contrast, irrational herding is driven by market sentiment  \citep{Banerjee1992}, fad \citep{Bikhchandani1992}, informational cascades \citep{BSV1998}, `word-of-mouth' effects from social imitation or influence \citep{Shiller2000, Hong2005} or irrational positive feedback trading from extrapolation of past growth rate \citep{Lakonishok1994, Nofsinger1999}. 

The challenge of diagnosing bubbles can thus be reduced to the detection and characterization of
the regularities associated with herding effects in real time, with the goal of predicting the potential 
upcoming regime-shift and possible large sell-off resulting from their evolution. Most existing
analyses have emphasized {\itshape herding of the overall market\/}, considering that investors
tend to synchronize their behavior across the whole investment horizon. Accordingly, 
methods to detect bubbles have been focused mostly on 
representations of the whole market performance, in general by using market indices. 
The rationale is that, during bubble regimes when widespread speculative behavior is prevalent, 
individual stocks tend to become cross-sectionally more tightly correlated in their behavior and follow
the general market dynamics. Notwithstanding this fact, the focus on market indices rather
than the constituting individual stocks is bound to overlook endogenous structures of {\itshape herding 
within the universe of stocks} (see e.g. \citet{Sias2004}, \citet{Choi2009}) and could
potentially miss useful patterns associated with the dynamics of speculation during the bubble build up. 
In particular, the financial crisis of 2008 that followed a large bubble regime \citep{Brunnermeier2012,SorCauwels2014}
suggests that there is important information for the development of systemic risk metrics imbedded
in the study of speculative bubble behavior in disaggregated industrial or firm level.

This paper presents an extension of more conventional bubble diagnosing methods by breaking down the structure of investment herding into its individual firm components. For this, we introduce the novel concept of the 
{\itshape Speculative Influence Network} (SIN), defined as a directional weighted network 
representing the causal speculative influencing relationships between all pairs of investment targets.
In other words, we quantify how speculative trading in one asset may draw speculative trading in other asset. 
Here, we will focus on stocks, but the method is more generally applicable to any basket of assets.
Specifically, we first estimate in real time the probability of speculative trading in each stock
in the basket of interest, using a bubble identifying technique that was introduced for whole market indices
but that we now extend to individual stocks. The strength of the bubble at the level of a single stock may increase or fade
multiple times during the development of a global market bubble, perhaps due to particular idiosyncratic properties
of the firm and of the corresponding industrial sector.  The {\itshape Hidden Markov Modeling} (HMM) approach, which is specially designed to calibrate regime-switching processes, is thus a convenient methodology to make
our bubble detection approach at the individual stock level more robust. Once we have characterized
the subset of stocks being qualified in a bubble regime, we calculate the 
{\itshape Transfer Entropy\/} (TE), which is a measure developed in Information Science
to quantify the casual relationship between variables. The underlying intuition
behind this method is that, if a stock $Y$ exhibits a strong speculative influence on another stock $X$, then 
the existence of speculative trading of $X$ can be predicated on the evidence of speculative 
trading on $Y$. In the language of Information Science, there is an information flow from $Y$ towards $X$,
which is quantified by the entropy transfer from $Y$ to $X$.   

The essential first step is of course to detect the presence of a bubble. For this, we 
borrow from the simple and generic prescription that a bubble is a transient regime characterized by 
 faster-than-exponential (or super-exponential) growth
(see e.g. \citep{HuslerHom13,KaizojiLeiss15,LeissNaxSor2015,SorCauwels2015,ArdialForroSor15} for recent empirical tests
and models). 
The  ``super-exponential'' behavior results from the existence of positive nonlinear feedback mechanisms, for
instance of past price increases on future returns rises \citep{HuslerHom13}.  
Searching for transient super-exponential price behavior avoids the curse of 
other bubble detection methods that rely on the need to estimate a fundamental value in order
to identify an abnormal pricing, the difference between the observed price and the supposed
fundamental value being attributed to the bubble component. This avoids the critique of 
\cite{Gurkaynak2008} in his study of the literature on rational bubbles 
and of econometric approaches applied to bubble detection, in which he reported that time-varying or regime-switching fundamentals could always be invoked to rationalize any declared bubble phenomenon. 
The second merit of super-exponential models is that they involve a mathematical formulation 
that inscribes the information on the end of the bubble, in the form of the critical time $t_c$
at which the model becomes singular in the form of an
a hyperbolic finite-time singularity (FTS). Actually, the bubble can end or burst before $t_c$, 
since, in the rational expectation bubble framework combined with the FTS models, $t_c$ is 
just the most probable time of the bubble burst but not the only one because various effects
can destabilize the price dynamics as time approaches $t_c$ (such as drying of liquidity).

The existing literature that has contributed until now to the concept of 
super-exponential bubble can be divided into two classes: (a) bubbles with deterministic 
maximum termination time $t_c$ and (b) bubbles with stochastic maximum termination time $t_c$. 
The first class takes its roots in the Johansen-Ledoit-Sornette (JLS) model \citep{JLS1, JLS2}. 
Developed within the rational expectation framework, the JLS model translates the aggregate speculative behavior
of investors into the existence of a critical dynamics associated with the self-organization of the 
system of mutually influencing traders. A parsimonious mathematical representation takes
the form of the  log-periodic power law singularity (LPPLS) dynamics. In addition to prices
exhibiting a super-exponential acceleration, there is a long-term volatility that accelerates according
to log-periodic oscillations embodying an accelerating cascade of exuberance followed
by limited panics resuming into exuberance and so on. In this framework, the critical time $t_c$
of the maximum duration of the bubble is also the 
most probable market crash time if it occurs. This critical time is treated as an intrinsic deterministic parameter 
in LPPLS bubble models. Since the introduction of the JLS model, there has been 
plenty of theoretical development of the LPPLS bubble framework, 
in particular expanding on the underlying mechanisms \citep{Didier1998,JS1998b,Ide2002} 
and on model calibration \citep{zhou2006, FiliSor2013, Lin2014}. Concurrently, the LPPLS literature 
has developed empirically with both post-mortem studies of past bubbles as well as real-time ex-ante
successful diagnostics of bubbles in a variety of financial markets, including western stock markets \citep{JLS1, JS2000b, Didier2006}, emerging stock markets \citep{JS2001b, giordano2006, zhou2009, cajueiro2009, Jiang2010}, real-estate markets \citep{zhou2006b, zhou2008}, oil markets \citep{Didier2009c}, forex markets \citep{JLS1, matsushita2006} and so on. 

Compared with LPPLS models, the other class of super-exponential bubble models, which explicitly consider the critical time $t_c$ as an endogenous random variable depending on the market state and the investors' aggregate expectations, have
been much less explored. The Sornette-Andersen super-exponential bubble model serves as the first attempt in this field \citep{AS1, AS2}. Under specific prescriptions of the nonlinear positive feedback effects, these 
bubble models have been proved to have their random critical times being distributed 
according to the inverse Gaussian distribution. Thereafter, the Sornette-Andersen model has been extended to 
richer situations in which the critical time satisfies an Ornstein-Uhlenbeck stochastic process \citep{SSCT}. 

In this paper, we propose, in the first step of our investigations, to focus on the Sornette-Andersen super-exponential bubble model as the primary prescription to develop diagnostics of speculative trading, before developing in a second step the analysis
of the structure of mutual interactions between stocks.
First, the Sornette-Andersen model has an elegant analytical solution and enjoys a more straightforward representation of speculative trading. Second, it is more tractable within an integrated stochastic framework and allows for a natural
and transparent implementation of the HMM approach, in which additional 
characteristics of speculation can be added. A first addition can be the condition of the stationarity of the critical end 
time due to the lack of synchronization. A second addition is the ingredient of log-periodicity of the long term volatility due
to the interplay between the inertia of transforming information into decision in the presence
of nonlinear momentum and price-reversal trading styles \citep{Ide2002}.

In order to demonstrate the usefulness of our proposed speculative influence network (SIN), we perform
our empirical study on the Chinese stock market. As an emerging financial market, the Chinese market 
has exhibited a number of dramatic bubbles and crashes \citep{Jiang2010,SorDemosZhang15}, reflecting strong
herding activities developing in a more opaque information environment 
with less regulatory control than mature western markets \citep{piotroski2011}. A few previous works have revealed
the existence of powerful speculative investors manipulating the market with relative facility, like in the so-called ``pump and dump'' strategy \citep{Khwaja2005, Zhou2005, He2009}. Additionally, less-informed small and medium size Chinese investors 
have been shown to be prone to trend-chasing speculation \citep{Li2009}. Moreover, empirical studies suggest an acute tendency of herding among Chinese investors \citep{Tan2008, Chiang2010}, especially for institutional investors whose herding is found to play a role in destabilizing the whole market \citep{Xu2013}. From mid-2005, the two major stock indices of the Chinese stock market, i.e., the Shanghai Stock Exchange Composite (SSEC) index and the Shenzhen Stock Component (SZSC) index, rose six-fold in just two years.  This extreme growth was then followed by a dramatic drop of about two-third of their peak values over a year. Though the transient prosperity of the financial market has been fueled by many compelling growth stories 
concerning the rapid fundamental growth of the Chinese economy, the roller-coaster performance of the whole market over 
the year of 2008 in particular offered hindsights about the existence of a bubble with a mixture of herding, over-optimistic speculation and positive feedback trading \citep{Jiang2010, Lin2014}. 

Our present study focuses on the Chinese stock market episodes from 2006 to 2008. To test the predictability of large cumulative losses associated with financial sell-offs in the Chinese stock market,  
we construct the SIN from 2006 to the end of 2007 when the market was still in a bullish state.
We then search for possible early-warning signals in the percentage maximum loss (\%{MaxLoss}) of each stock during 2008,
using measures derived from the network-based analysis. The network is constructed by using the representative stocks portfolios and indices of various industrial sectors. Also, considering the implications to measure systemic risks, we expand the
single node representing the  financial sector in the original network to become a full 
sub-network made of all stocks from the financial sector, including banks, trust companies, securities firms and insurance companies.  
Our main empirical findings is that the total net influence effect measured with the Transfer Entropy (TE) method
in industrial sectors (except for the financial sector and for financial institutions) is a significant determinant
of the \%{MaxLoss} of a number of stocks.  Moreover, the gross TE to all industrial sectors significantly explains the \%{MaxLoss} of financial institutions, whereas the total TE to or from financial institutions are not good explanatory indicators to predict the \%{MaxLoss} of non-financial sectors.
These results suggest that the SIN can not only be used to construct a cross-sectional anatomy of bubbles but can also help in complementing the analytics of linkage-based systemic risk measures, by estimating the interconnection of institutions with
respect to their vulnerability to exuberant speculation.  

The structure of this article is as follows. Section \ref{S2} first introduces the Sornette-Andersen super-exponential bubble model with stochastic termination times and then discusses the detectability of speculative trading using the HMM approach from the regime-switching perspective of bubble onset and burst. In Section \ref{S3}, we investigate how to build the speculative influence network (SIN) for a financial market with the help of the Transfer Entropy between stocks. Section \ref{S4} studies the practical relevance  of the SIN and also explores possible measures of systemic risks that can 
be derived from it. An empirical application is made to the Chinese stock market. Finally, we conclude and sketch out potential extensions in Section \ref{S5}.

\section{Regime-Switching speculative bubbles with super\\ exponential growth}\label{S2}

\subsection{The Sornette-Andersen bubble model}\label{S21}

According to the principle underlying the whole analysis of this paper, speculative trading 
during a bubble is characterized by positive feedbacks, i.e., the higher the number of interested investors and the higher the price, the larger the increase of new stock purchases and the higher the price growth. 
Positive feedbacks are contributed by a large diversity of herding activities, leading to copycats rushing to follow their leaders \citep{chincarini2012}. It is worth noting that, while herding has been largely documented to be a trait of noise traders, it is actually rational for bounded rational agents to also enter into social imitation, because the collective behavior may reveal information otherwise hidden to the agents. Moreover, 
even in the absence of information, it may also be rational to imitate one's social network because it may reflect
the consensus who decisions are incorporated in subsequent returns \citep{Didier2000}.

Once positive feedback takes over, the financial market, like all systems with positive feedback, enters a state of increasing unbalance, with unsustainable increasing rates of return. Positive feedbacks lead to 
faster-than-exponential price appreciation: as prices increase, the expectation of future growth increases even further, pushing anticipation of future returns upwards. This means that the growth rate grows itself. As a constant growth rate of the price (namely, a constant return) corresponds to an exponential price trajectory, a growing growth rate leads (in many specifications) to a faster-than-exponential price path with a finite-time singularity signaling the end of the bubble. The technical analysis literature casually refers to this pattern as ``parabolic'' or ``hyperbolic'' growth
because of the corresponding upward curvature (convexity) of the log-price as a function of time.
The important point for our purpose is that the transient faster-than-exponential price path provides a diagnostic of speculative bubbles that is fundamentally different from the standard academic methods
that emphasize more the detection of just abnormal exponential growth regimes above the fundamental price or of mildly explosive regimes.
One could term the regimes we refer to as being ``super-explosive'', to emphasize the difference with the term ``explosive'' 
often attributed to an exponential regime (which is the norm in finance and in economics, as well as in demographics, because
an exponential growth just reflects the mechanism of proportional growth, which is nothing but compounding interest in finance).

In the context of speculative bubbles, the Sornette-Andersen model is arguably the most parsimonious representation to capture a transient ``super-explosive'' signature in a continuous stochastic framework, which formulates the interplay between multiplicative noise and nonlinear positive feedbacks onto future returns and volatility. According to the model, the price dynamics in the bubble regime is assumed to be described by the following stochastic differential equation \citep{AS1}:
\begin{equation}\label{E1}
\frac{d p}{p}=[\,\mu_1(p)+\mu_2(p)\,] dt+\sigma(p)d W -\kappa dj
\end{equation}
where the first drift term $\mu_1(p)=\mu p^n, n>0$ represents the nonlinear positive feedback effect that larger asset price feeds a larger growth rate \citep{HuslerHom13}, which leads to even larger price and so on. The parameters $\mu$ and $n$ are respectively the effective drift coefficient and strength of nonlinear positive feedback effects. The diffusion part 
is specified by $\sigma(p)=\sigma p^n, n>0$, with the same strength of the nonlinear positive feedbacks of price, 
and can be interpreted as a conditional nonlinear heteroskedasticity. Intuitively, the term may be 
rationalized by the prevalence of hedging strategies using general Black-Scholes option models and of other herding behaviors discussed before. The parameter $\sigma$ denotes the magnitude of the stochastic component, which sets the scale of volatility. 
The last term $-\kappa dj$ represents the crash, where $dj=1$ when the crash occurs and $dj=0$ otherwise,
and $\kappa$ is the crash amplitude. The crash hazard rate $h(t) :=  {\rm E}[dj]$ (where ${\rm E}[.]$ is the expectation operator)
is specified by the condition of no arbitrage, which translates into the fact that the price is a 
Martingale: ${\rm E}[\frac{d p}{p}]=0$. This leads to
\begin{equation}
h(t) = {1 \over \kappa} (\mu_1(p)+\mu_2(p))~.
\label{strhtyhgv}
\end{equation}

Without the drift term $\mu_2(p)$ and neglecting the crashes, expression (\ref{E1}) transforms into the standard Stratonovich 
deterministic equation,
\begin{equation}
\frac{dp}{p}=(\mu dt+ \sigma dW) p^n, \quad n>0
\end{equation}
giving the intuitive interpretation that the market noise $dW$ gradually accumulates in the price through a simple multiplicative mechanism
reinforced by the nonlinear term $p^n$.

The second drift term is set to be $\displaystyle\mu_2(p)=\frac{n+1}{2} \sigma^2(p)$. As the market required a risk premium for the existing volatility and upcoming crash, this term serves to balance the subtle interplay between multiplicative noise and nonlinear positive feedbacks. With this additional convenient device,  It\^{o} calculation of these stochastic differential equations \eqref{E1} can be drastically simplified, as we will see soon. 

The most striking feature of the model is that the nonlinearity ($n>0$) creates a singularity in finite time and the multiplicative noise makes it stochastic. To see this, let us first consider the case $\sigma \to 0$ and $\kappa \to 0$, for which eq.\eqref{E1} reduces to the deterministic price dynamics  $\dfrac{d p}{p}=\mu p^n dt$, whose solution reads
\begin{equation}
p(t)=(n \mu)^{-\frac{1}{n}}[\,t_c-t\,]^{-\frac{1}{n}},\quad t_c=\frac{p_0^{-n}}{n\mu}
\label{trhetuyju4w}
\end{equation}
where $p_0=p(t=t_0)$. Note that such deterministic dynamics is characterized by a fixed critical time, $t_c$, 
such that $p(t)$ diverges when time approaches $t_c$ from below. This critical time is called 
a movable singularity, because it depends on the initial condition $p_0$ in addition to the parameters of the model.
Reintroducing a non-zero volatility, and conditional on the absence of a crash ($dj=0$),
 It\^{o} calculus applied to \eqref{E1} provides the following
explicit analytical solution (See \cite{AS1} for the complete derivation therein):. 
\begin{equation}\label{E2}
p(t)=[\,n\mu(t_c-t)-n\sigma W_t\,]^{-\frac{1}{n}}, \quad t_c=\frac{p_0^{-n}}{n\mu}
\end{equation}
This solution recovers, as it should, the deterministic solution (\ref{trhetuyju4w}) for $\sigma =0$.
The price now exhibits a stochastic behavior, which can still diverge when there is a time such that
the denominator $n\mu(t_c-t)-n\sigma W_t$ crosses $0$. At $t=0$, the denominator 
starts from an initial positive value $n\mu t_c$. In the presence of the negative drift $-n\mu t$ and notwithstanding
the presence of the Wiener process, the denominator will cross $0$ with probability $1$.
The corresponding stochastic critical time $\tilde{t}_c$ at which the price diverges is thus controlled by a first-passage problem
when the drifting random walk motion $n\mu t+n\sigma W_t$ first encounters the value $n\mu t_c =p_0^{-n}$. 
From standard results of first-passage problems \citep{Redner01,JeanblancYor09}, the stochastic 
critical time $\tilde{t}_c$ is found to be
distributed according to the {\itshape Inverse Gaussian} distribution
\begin{equation}
\tilde{t}_c \sim \mathcal{IG} \left(\,\frac{p^{-n}_0}{n\mu},\, \left[\frac{p^{-n}_0}{n\sigma}\right]^2\,\right)~.
\end{equation}

In fact, with the specification (\ref{E1}) augmented by (\ref{strhtyhgv}), the price never diverges. This is because,
as the price accelerates, so does even more its instantaneous growth rate $\mu_1(p)+\mu_2(p)$
and therefore so does the crash hazard rate via expression (\ref{strhtyhgv}). This is aimed
at embodying the fact that the bubble collapses as a result of the ebb of speculation and withdrawal of 
intended demand when an unsustainable state is becoming more obvious or due to increasing
scarcity of money and credit. In other words, 
a crash always occurs before the price goes too high, ensuring long-term stationarity
\citep{AS1}.

Note that the price dynamics (\ref{E2}) reduces to the geometric random walk when the strength of positive feedback 
vanishes ($n=0$):
\begin{equation}
\lim_{n \to 0} p(t)=p_0\cdot\lim_{n \to 0} [\,1-n (\mu t+\sigma W_t)\,]^{-\frac{1}{n}}=p_0 \, \mathrm{e}^{\,\mu t+\sigma W_t} ~,
\end{equation}
which is indeed the solution of the standard Black-Scholes framework obtained as the limit $n \to 0$ of 
equation (\ref{E1})  (with $\kappa=0$).

\subsection{Hidden Markov modelling (HMM) calibration approach}\label{S22}

We propose to extend the Sornette-Andersen model summarized in the previous section 
by combining it with a standard geometric random walk, supposed to represent the normal regime of markets.
The complete price process is then described by regime switching between the dynamics described by 
the Sornette-Andersen model with significantly positive value of $n$ and the degenerated geometric random walk for $n=0$. 
Then, the existence of speculative trading can be judged by estimating the probability that the 
market is in bubble state ($n>0$) or in the normal state ($n=0$). For this, we convert the bubble model into a Hidden Markov modeling (HMM) framework and then calibrate all parameters in the model to identify the onsets and ends of bubbles,  
following a procedure that we now describe. Because crashes represented by the jump term $dj$ in 
(\ref{E1}) as rare, we chose to not identify them and only rely on the structural differences between
the process conditional on no crash for $n>0$ and $n=0$.

For the sake of conciseness and efficiency of the presentation, in the following discussion, we 
introduce a novel mathematical representation of probability distribution functions and of conditional density functions. 
They resemble the famous ``bra-ket'' notations in Dirac  notations in Quantum Mechanics. To be clear, 
no new results are derived, it is just a notation used for its compactness and coherence.
The notations are as follows. We use ``bra" to denote the probability density for a random variable, i.e., $\p{x}:= f_{\tilde{x}}(x)$; We use ``ket" to denote the ``conditional on" operation. Formal ``multiplication'' of a bra-vector and  a ket vector 
gives the conditional density, i.e., 
$\p{x}\cdot\cd{y}=\np{x}{y}:=f_{\tilde{x}\mid\tilde{y}}(x | y)$. In the Appendix \ref{AP1},
we show that, with the addition of a few formal transformation rules, more complicated expressions can be nicely 
represented, such as the decomposed conditional joint density or Bayesian formulas for multivariate conditional densities.
This is convenient to present the derivation of the formulas of our proposed HMM procedure developed
to estimate the super-exponential Sornette-Andersen bubble model from the perspective of a
state-dependent regime-switching process.          

In order to formulate the HMM, we assume there is an underlying stochastic state process  $s_t$  that
indicates the prevalence of speculative trading, which is unobserved by the investors. The state is switching between $0$ and $1$, signalling respectively the ``normal" and bubble states. We use 
\begin{equation}
q_{ij}=\np{s_t=j}{s_{t-1}=i}~,~~~~~~ i,j=0,1
\label{wynjht}
\end{equation}
to denote the element $(i,j)$ of the Markov-chain transition probability matrix. 

Two cases must be considered in the absence of switching, whose specifications derive straightforwardly from
the definition of the Sornette-Andersen model:
\begin{itemize}
\item $(s_t,s_{t-1})=(0,0)$: The market remains in the state without bubble. Therefore, the price dynamics $p_t$ follows the geometric Brownian motion, i.e., $y_t\mid y_{t-1}\,\sim\, \mathcal{N}(\mu_0+y_{t-1},\sigma_0^2)$, where $y_t=\ln p_t$. 
Specifically, the transition probability for $y_t$ is expressed as 
\begin{equation}
\ln\np{y_t}{s_t=0,s_{t-1}=0,y_{t-1}}=-\dfrac{1}{2}\ln 2\pi-\ln\sigma_0-\dfrac12\cdot\left(\dfrac{y_t-y_{t-1}-\mu_0}{\sigma_0}\right)^2~.
\end{equation}
\item $(s_t,s_{t-1})=(1,1)$: The market is experiencing a continuing super-exponential bubble 
characterized by the prevalence of speculation and herding associated with positive feedback. Hence, the price dynamics $p_t$ is captured by the Sornette-Andersen model summarized by eq.\eqref{E2} in the absence of crash. The 
corresponding discrete formulation, $p_t^{-n}\mid p_{t-1}^{-n}\,\sim\,\mathcal{N}(p_{t-1}^{-n}-n\mu_1,n^2\sigma_1^2)$ leads to the transition probability for $y_t$ (See Appendix \ref{AP2} for the detailed proof) given by
\begin{equation}\label{E:tpfb}
\ln\np{y_t}{s_t=1,s_{t-1}=1,y_{t-1}}=-\dfrac{1}{2}\ln 2\pi-\ln n\sigma_1-\dfrac12\cdot\left(\dfrac{e^{-ny_t}-e^{-ny_{t-1}}+n\mu_1}{n\sigma_1}\right)^2+\ln n -n y_t~.
\end{equation} 
\end{itemize}

In the presence of a regime switch, we have to make some additional assumption and we choose the simplest possible one:
\begin{numcases}{}
\np{y_t}{s_t=0,s_{t-1}=1,y_{t-1}}= \biggl|\dfrac{1}{\mu_0}\biggr|\cdot\mathbf{1}_{\{-\kappa \leq y_t<y_{t-1}\}}  \label{mjyrgds}\\
\np{y_t}{s_t=1,s_{t-1}=0,y_{t-1}}=\biggl|\dfrac{1}{\mu_1}\biggr|\cdot\mathbf{1}_{\{ y_{t-1} \leq y_t \leq \kappa \}},  \label{yjyssa}
\end{numcases}
where $\mathbf{1}_{\{\text{condition}\}}$ is the indicator function that equals to $1$ when condition is met, otherwise it gives $0$. 
Expression (\ref{mjyrgds}) writes that, at the time when a bubble ends, the price drops 
by an amount that is bounded by the crash amplitude $\kappa$. Similarly, expression (\ref{yjyssa}) expresses that, a bubble starts, the price tends to appreciate. We bound the maximum amplitude of appreciation by the scale set by $\kappa$.
The simplicity of (\ref{mjyrgds}) and (\ref{yjyssa}) allows for an analytical expression of the likelihood function, which 
is convenient for its maximisation and determination of the model parameters, as we shall seen below. 

The price dynamics of our state-dependent Markov-switching periodically collapsing bubble model is entirely encoded in the vector of parameters $\pmb{\theta}=(\mu_0,\sigma_0, \mu_1, \sigma_1, n, q_{00}, q_{11}, q_{01}, q_{10})$. Because the 
underlying state variable $s_t$ is hidden, it is natural to apply the 
{\itshape Expectation-Maximumization (EM)} algorithm to estimate $\pmb{\theta}$ \citep{Dempster1977,Kim1999}.
In a nutshell, the EM algorithm is implemented by iteratively performing two steps until meeting the convergence condition : (1) calculation of the expectation of the log-likelihood evaluated using the current parameters estimates (called E-step) and (2)
maximization of the obtained averaged log-likelihood function to obtain the optimized posterior estimates (called M-step). 
Applied to our model, the E-step reads
\begin{equation}\label{E:estep}
\mathcal{\ln L}_{\pmb{\theta}\,\mid\,\pmb{\theta}^{(k-1)}}=\sum_{1\cdots T} \left(\ln\p{\y{T},\s{T}}_{\pmb{\theta}}\right)
\,\,\p{\y{T},\s{T}}~_{\pmb{\theta}^{(k-1)}}
\end{equation}
where the notation $\cdot^{T}_{\multimap}$ means that the corresponding variable is multidimensional
and covers all the time steps from the start $0$ to the end $T$ of the time window.
The symbol $\displaystyle \sum_{1\cdots T}$ means that the summation over all states $s_t, t=1,\cdots,T$ 
is performed. The above expectation step (\ref{E:estep}) can be expanded into (See Appendix \ref{AP3.1} for the detailed derivation)
\begin{equation}\label{E:estep1}
\mathcal{\ln L}_{\pmb{\theta}^{(k)}\,\mid\,\pmb{\theta}^{(k-1)}}=
\p{\y{T}}_{\pmb{\theta}^{(k-1)}}\cdot\left(\sum_{t=1}^T \sum_{\substack{\;\;s_t=1,0\\ \!\!s_{t-1}=1,0}}\Bigl(\ln \np{y_{t}}{s_t,s_{t-1},y_{t-1}}_{\pmb{\theta}^{(k)}}+\ln\np{s_t}{s_{t-1}}_{\pmb{\theta}^{(k)}}\,\Bigr)\np{s_{t-1},s_t}{\y{T}}_{\pmb{\theta}^{(k-1)}}\right)
\end{equation} 

In the M-step (See Appendix \ref{AP3.2}, \ref{AP3.3} and \ref{AP3.4}), 
we then obtain the following expressions for the corresponding posterior estimates of the parameters:
\begin{itemize}
\item $\mu_0^{(k)}=\dfrac{\sum_{t=1}^T \omega_{(0,0);t}^{(k-1)}(y_t-y_{t-1})}{\sum_{t=1}^T\omega_{(0,0);t}^{(k-1)}},\qquad \omega_{(0,0);t}^{(k-1)}:=\np{s_t=0,s_{t-1}=0}{\y{T}}_{\pmb{\theta}^{(k-1)}}$
\item $\sigma_0^{(k)}=\sqrt{\dfrac{\sum_{t=1}^T \omega_{(0,0);t}^{(k-1)}(y_t-y_{t-1}-\mu_0^{(k)})^2}{\sum_{t=1}^T\omega_{(0,0);t}^{(k-1)}}},\qquad \omega_{(0,0);t}^{(k-1)}:=\np{s_t=0,s_{t-1}=0}{\y{T}}_{\pmb{\theta}^{(k-1)}}$
\item $\mu_1^{(k)}=\dfrac{\sum_{t=1}^T \omega_{(1,1);t}^{(k-1)}(p_{t-1}^{-n}-p_t^{-n})}{n \sum_{t=1}^T\omega_{(1,1);t}^{(k-1)}},\qquad \omega_{(1,1);t}^{(k-1)}:=\np{s_t=1,s_{t-1}=1}{\y{T}}_{\pmb{\theta}^{(k-1)}}$
\item $\sigma_1^{(k)}=\sqrt{\dfrac{\sum_{t=1}^T \omega_{(1,1);t}^{(k-1)}(p_{t}^{-n}-p_{t-1}^{-n}+n\mu_1^{(k)})^2}{n^2 \sum_{t=1}^T\omega_{(1,1);t}^{(k-1)}}},\qquad \omega_{(1,1);t}^{(k-1)}:=\np{s_t=1,s_{t-1}=1}{\y{T}}_{\pmb{\theta}^{(k-1)}}$
\item The value $n^{(k)}$ is solution of the following nonlinear univariate equation 
\begin{equation}\label{E:solven}
\sum_{t=1}^T \left(-\dfrac{(p_t^{-n}-p_{t-1}^{-n}+n\mu_1)(-p_t^{-n}\ln p_t+p_{t-1}^{-n}\ln p_{t-1}+\mu_1)}{n^2\sigma_1^2}+\frac{1}{n}-\ln p_t\right)\,\omega_{(1,1):t}^{(k-1)}=0
\end{equation}
\item $q_{ij}^{(k)}=\dfrac{\sum_{t=1}^T\np{s_t=j,s_{t-1}=i}{\y{T}}_{\pmb{\theta}^{(k-1)}}}{\sum_{t=1}^T\np{s_{t-1}=i}{\y{T}}_{\pmb{\theta}^{(k-1)}}},\quad i,j=0,1$
\end{itemize}

In order to use these expressions, we need to have the values of  $\omega_{(0,0);t}^{(k-1)}$ and $\omega_{(1,1);t}^{(k-1)}$,
which requires to determine $\np{s_t,s_{t-1}}{\y{T}}$ and $\np{s_t}{\y{T}}$. For this, we first adopt 
a Hamilton filtering procedure (see \citep{Hamilton1989}) to obtain $\np{s_{t}}{\y{t}}, 0\le t\le T$, which 
can be decomposed in three steps:
\begin{enumerate}
\item {\itshape  Prediction Step}: from $\np{s_{t-1}}{\y{t-1}}$ to $\np{s_t,s_{t-1}}{\y{t-1}}$ 
\begin{align}
\np{s_t,s_{t-1}}{\y{t-1}}&=\p{s_t,s_{t-1}}\cdot\cd{\y{t-1}}
= (\np{s_t}{s_{t-1}}\p{s_{t-1}})\cdot\cd{\y{t-1}}=\np{s_t}{s_{t-1}, \y{t-1}}\np{s_{t-1}}{\y{t-1}}\notag\\
&=\np{s_t}{s_{t-1}}\np{s_{t-1}}{\y{t-1}}
\end{align}
This is nothing but the composed conditional joint density (See Appendix \ref{AP1}).
\item {\itshape Updating Step}: from $\np{s_t,s_{t-1}}{\y{t-1}}$ to $\np{s_t,s_{t-1}}{\y{t}}$
\begin{align}
\np{s_t,s_{t-1}}{\y{t}}&=\np{s_t,s_{t-1}}{\y{t-1},y_t}=\np{s_t,s_{t-1}}{\y{t-1}}\frac{\np{y_t}{s_t,s_{t-1},\y{t-1}}}{\np{y_t}{\y{t-1}}}\label{E:skkaln}\\
&=\dfrac{\np{y_t}{s_t,s_{t-1},\y{t-1}}\np{s_t,s_{t-1}}{\y{t-1}}}{\sum_{t-1,t}\np{y_t}{s_t,s_{t-1},\y{t-1}}\np{s_t,s_{t-1}}{\y{t-1}}}
\end{align}
The validity of eq. \eqref{E:skkaln} is based on the multivariate Bayesian formula (See Appendix \ref{AP1} for details). 
\item {\itshape Summation Step}: from $\np{s_t,s_{t-1}}{\y{t}}$ to $\np{s_{t}}{\y{t}}$
\begin{align}
\np{s_{t}}{\y{t}}&=\p{s_t}\cdot\left(\sum_{t-1}\cd{s_{t-1}}\p{s_{t-1}}\right)\cd{\y{t}}=\left(\sum_{t-1}\np{s_t}{s_{t-1}}\p{s_{t-1}}\right)\cd{\y{t}}=\left(\sum_{t-1}\p{s_t,s_{t-1}}\right)\cd{\y{t}}\notag\\
&=\sum_{t-1}\np{s_t,s_{t-1}}{\y{t}}
\end{align}
This provides the {\itshape filtering probability} $\np{s_{t}}{\y{t}}$, which quantifies the probability for a given bubble state to be present, 
based on the current available market information. 
\end{enumerate}

Second, with the obtained filtering probability, we take advantage of a backward smoothing approach to estimate $\np{s_t,s_{t-1}}{\y{T}}$ and $\np{s_t}{\y{T}}$, which is inspired by Kim's smoothing routine (See \cite{Kim1999} for a discussion therein). Let $\iy{t+1}$ denote all of the 
remaining log-prices $\ln(p_t)$ expressed at time following time $t$ in the observation window from $1$ to $T$, i.e., $\y{T}=\{\y{t},\iy{t+1}\}$. The backward prediction is then expressed as
\begin{align}
\np{s_{t+1},s_t}{\y{T}}&=\np{s_t}{s_{t+1},\y{T}}\np{s_{t+1}}{\y{T}}=\np{s_t}{s_{t+1},\y{t},\iy{t+1}}\np{s_{t+1}}{\y{T}}\label{E:dddlla}\\
&=\np{s_t}{s_{t+1},\y{t}}\np{s_{t+1}}{\y{T}}
=\np{s_t}{\y{t}}\frac{\np{s_{t+1}}{s_t,\y{t}}}{\np{s_{t+1}}{\y{t}}}\np{s_{t+1}}{\y{T}}\label{E:llnkaoo}\\
&=\np{s_t}{\y{t}}\np{s_{t+1}}{\y{T}}\frac{\np{s_{t+1}}{s_t}}{\sum_t\np{s_{t+1}}{s_t}\np{s_t}{\y{t}}}~.
\end{align}
The derivation of the r.h.s. in \eqref{E:llnkaoo} is based on the multivariate Bayesian formula (See Appendix \ref{AP1}). The detailed derivation from \eqref{E:dddlla} to \eqref{E:llnkaoo} can be found in Appendix \ref{AP3.5}. 
This finally leads to the following expression to calculate $\np{s_{t}}{\y{T}}$, which can be called {\itshape smoothing probability}:
\begin{equation}
\np{s_{t}}{\y{T}}=\sum_{t+1}\np{s_t}{s_{t+1}}\np{s_{t+1}}{\y{T}}=\sum_{t+1}\np{s_{t+1},s_t}{\y{T}},\quad t=1,\cdots, T-1~.
\label{tyjunhwg}
\end{equation}

As its mathematical denotation to indicate, the {\itshape smoothing probability} $\np{s_{t}}{\y{T}}$ 
measures the likelihood of the existence of bubbles in a given time interval by including the information of 
future prices. For our purpose of diagnosing a bubble and its possible dangerous time period for a burst, 
it is necessary to use the causal {\itshape filtering probability} $\np{s_{t}}{\y{t}}$ and not the non-causal 
{\itshape smoothing probability} $\np{s_{t}}{\y{T}}$. The later can however be useful 
for parameter estimations, which themselves impact on bubble detection.

\subsection{Empirical estimation of the bubble probability in 16 financial markets \label{wrntuyjnhwg}}

\begin{figure}[p]
\centering
\includegraphics[totalheight=21cm]{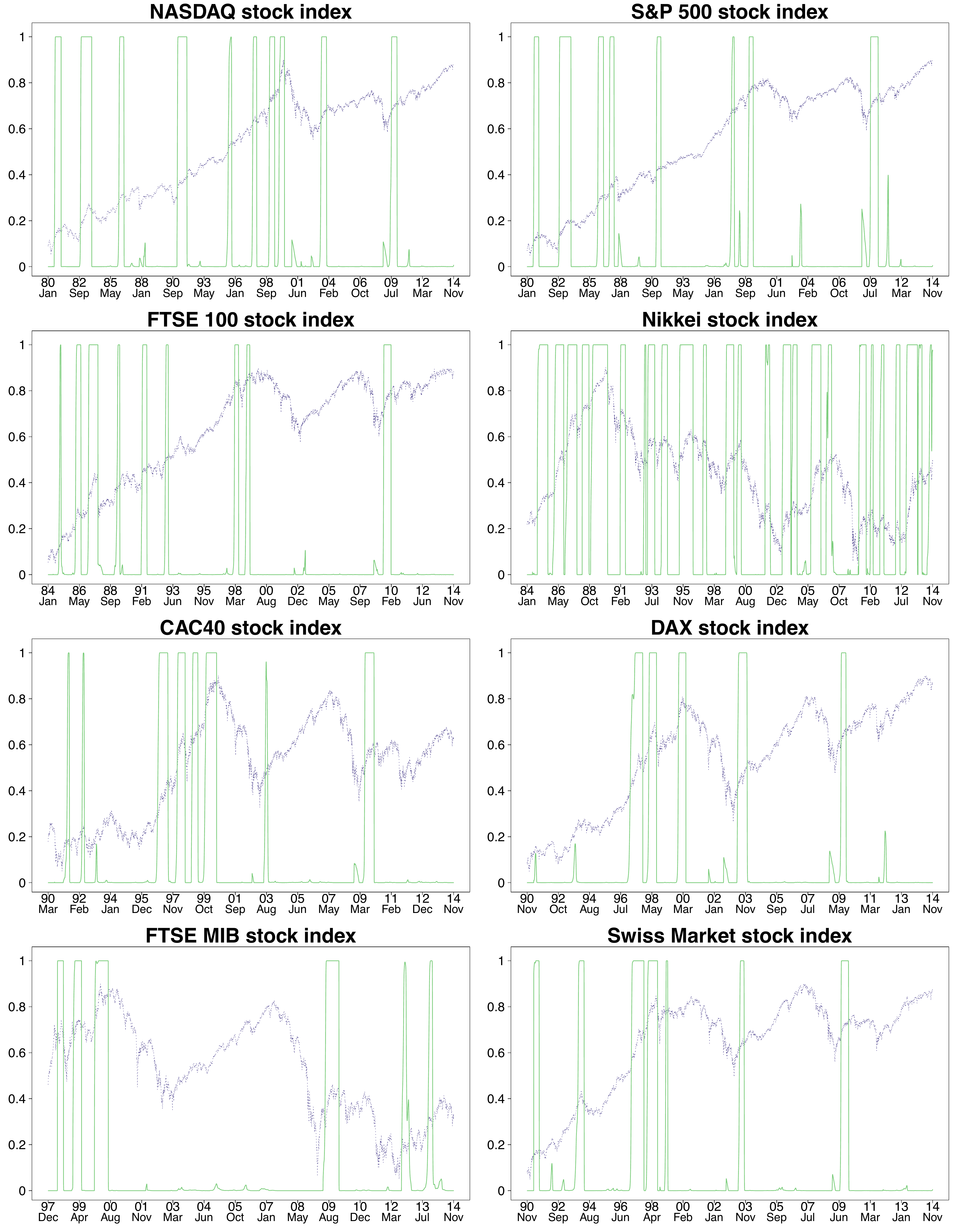}
\caption{\small Estimated daily smoothing probabilities $\np{s_{t}}{\y{T}}$ of the periodical collapsing super-exponential growth bubble embedded in the Hidden Markov Modeling (HMM) procedure represented with the  
palegreen thin continuous curve bouncing between $0$ and $1$ for 
eight stock exchanges: U.S. (Nasdaq and S\&P500), U.K., Japan, France, Germany, Italy and Switzerland. All indices are 
given in logarithmic scale and are rescaled into [0,1] for comparison. In order to mitigate the effects of identifying regime-shift too often as a result of daily price volatility, all indices are geometrically averaged over 100 days according to 
 $\ln p^{\rm filtered}_t=(1/100) \sum_{i=0}^{99} \ln p_{t-i}$, before implementing the EM algorithm. }
\label{F:1}
\end{figure}

\begin{figure}[p]
\centering
\includegraphics[totalheight=21cm]{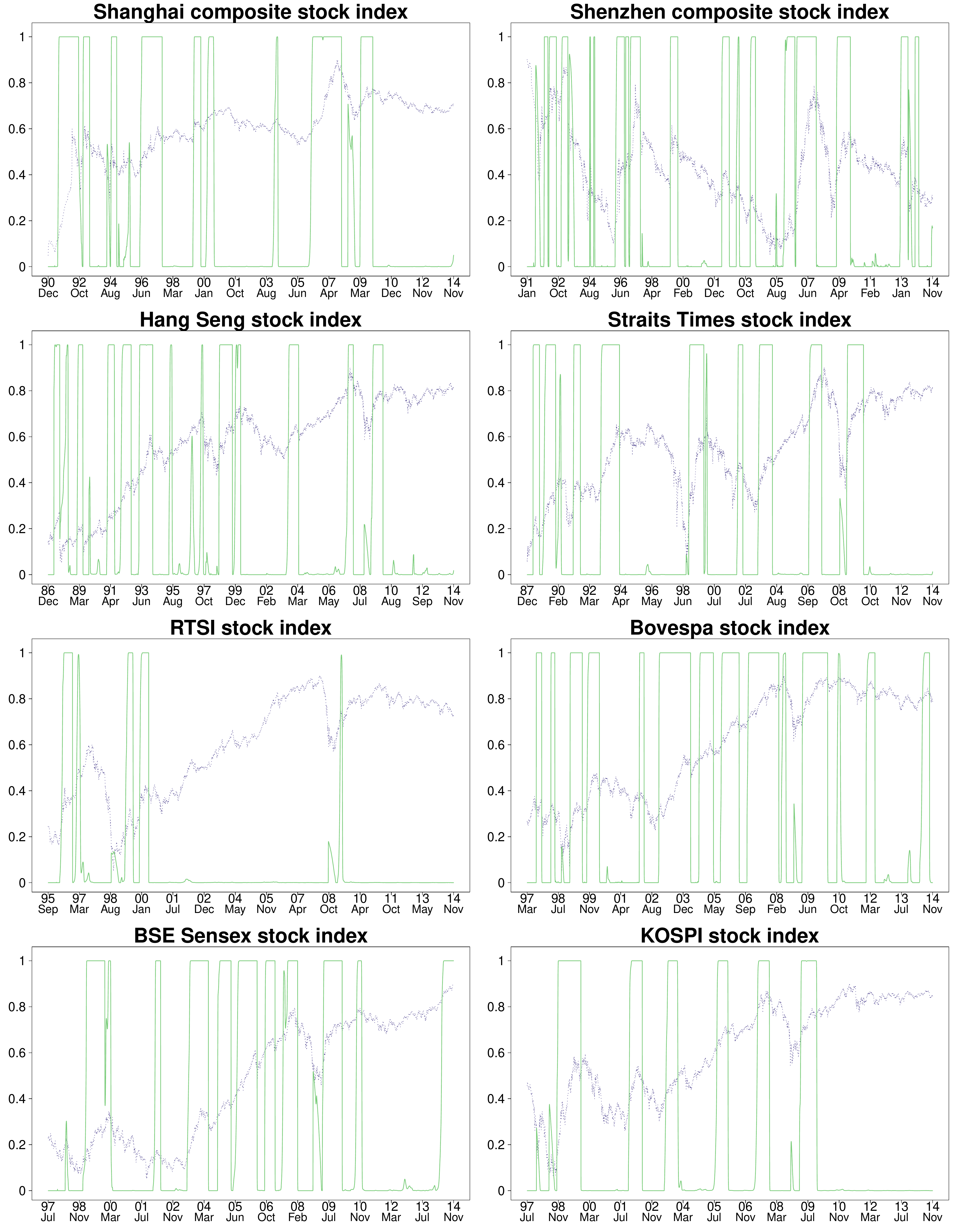}
\caption{\small same as figure \ref{F:1} for eight typical indices from stock exchanges of BRICs and other Asian markets, including China mainland, Hong Kong, Singapore, Russia, Brazil, India and Korea.
}\label{F:2}
\end{figure}

\begin{figure}[p]
\centering
\includegraphics[totalheight=21cm]{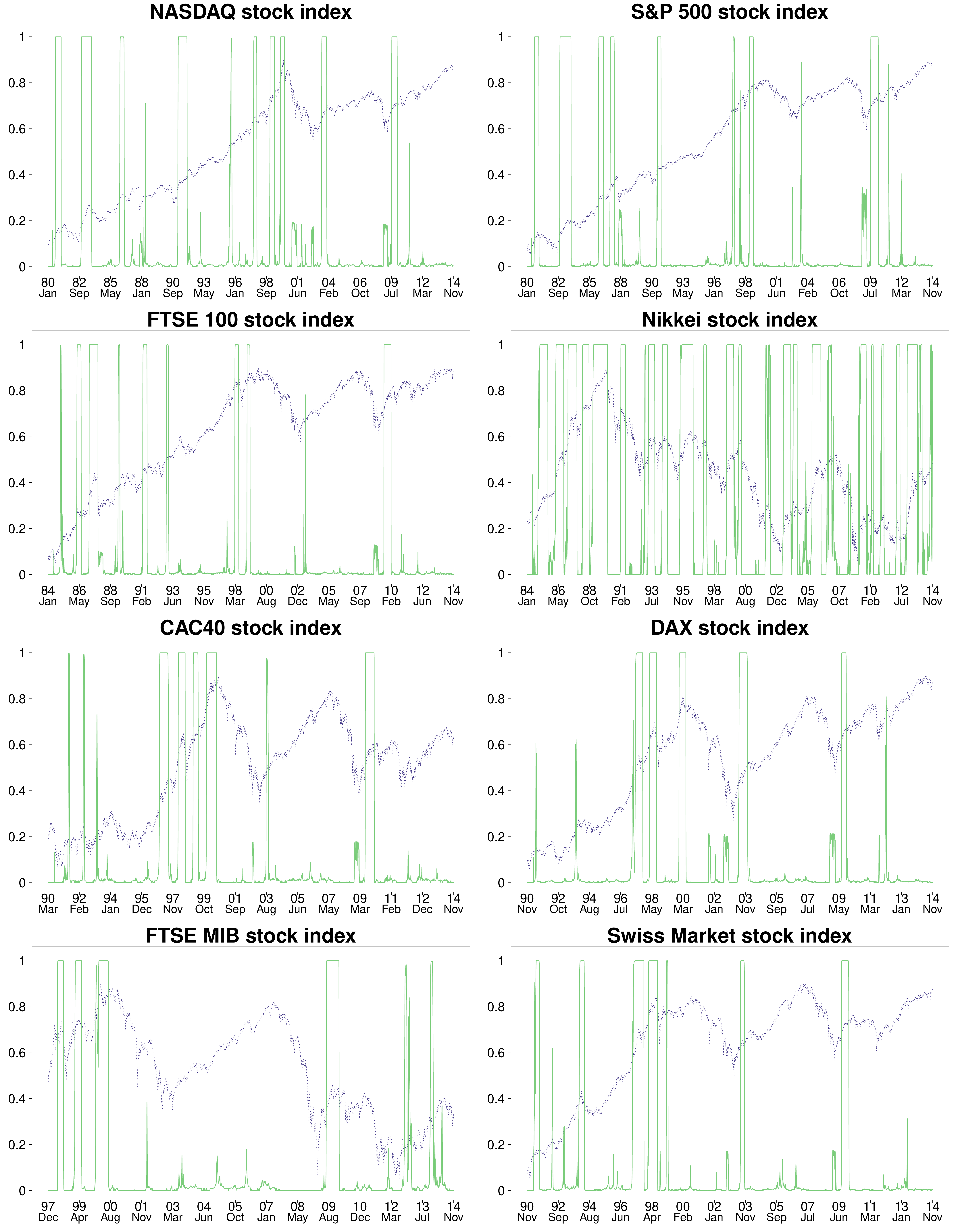}
\caption{\small Estimated daily filtering probabilities $\np{s_{t}}{\y{t}}$ of the periodical collapsing super-exponential growth bubble embedded in the Hidden Markov Modeling (HMM) procedure represented with the  
palegreen thin continuous curve bouncing between $0$ and $1$ for 
eight stock exchanges: U.S. (Nasdaq and S\&P500), U.K., Japan, France, Germany, Italy and Switzerland. All indices are 
given in logarithmic scale and are rescaled into [0,1] for comparison. In order to mitigate the effects of identifying regime-shift too often as a result of daily price volatility, all indices are geometrically averaged over 100 days according to 
 $\ln p^{\rm filtered}_t=(1/100) \sum_{i=0}^{99} \ln p_{t-i}$, before implementing the EM algorithm.}
 \label{F:1'}
\end{figure}

\begin{figure}[p]
\centering
\includegraphics[totalheight=21cm]{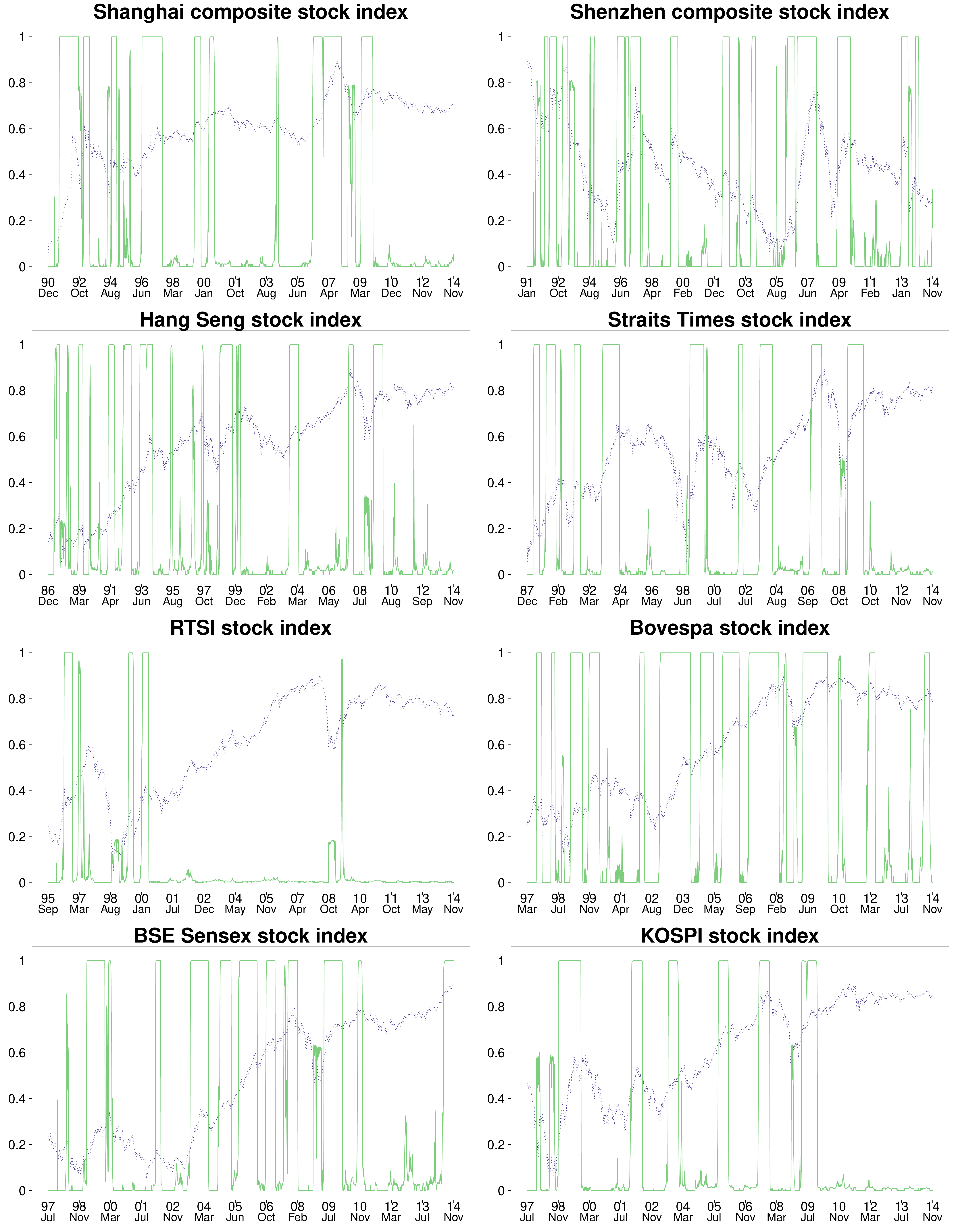}
\caption{\small same as figure \ref{F:1'} for eight typical indices from stock exchanges of BRICs and other Asian markets, including China mainland, Hong Kong, Singapore, Russia, Brazil, India and Korea.}
\label{F:2'}
\end{figure}

Figs. \ref{F:1} and \ref{F:2} show the estimated daily smoothing probability for the presence of super-explosive bubbles based on the HMM procedure in sixteen different global financial markets, represented by their major stock indices. Correspondingly, Figs. \ref{F:1'} and \ref{F:2'} display the estimated daily filtering probability for the presence of super-explosive bubbles for the same sixteen global financial markets. 
The studied stock indices include the US NASDAQ index and S\&{P} 500 stock index from 
the New York Stock Exchange, the FTSE 100 index from London Stock Exchange, the Nikkei 225 index from Tokyo Stock Exchange, the CAC40 stock index from  Euronext Paris, the DAX index from Frankfurt Stock Exchange in Germany, Switzerland's blue-chip stock market index (SMI), the FTSE MIB stock index from Borsa Italiana, the two major stock indices from Chinese stock market (Shanghai composite index and Shenzhen composite index), the Hang Seng index from Hong Kong Stock Exchange, the Straits Times stock index from Singapore Exchange, the Russia Trading System (RTS) index from Moscow Stock Echange, the Bovespa index from the S\~{a}o Paulo Stock, Mercantile \& Futures Exchange, the SENSEX index from Bombay Stock Exchange in India and the Korea Composite Stock Price Index(KOSPI). In the implementation of our EM procedure, the condition of convergence of the algorithm is specified as 
\begin{equation}
\Delta=\frac{\mathcal{\ln L}_{\pmb{\theta}^{(k+1)}\,\mid\,\pmb{\theta}^{(k)}}-\mathcal{\ln L}_{\pmb{\theta}^{(k)}\,\mid\,\pmb{\theta}^{(k-1)}}}{\mathcal{\ln L'}_{\pmb{\theta}^{(k)}\,\mid\,\pmb{\theta}^{(k-1)}}}\le 0.0001
\end{equation} 

\noindent Several observations are worth describing. 

$\bullet$ There is an excellent correspondence between the
estimated daily smoothing and filtering probabilities in essentially all cases: the periods identified in the super-exponential
bubble state with the {\itshape smoothing probability} $\np{s_{t}}{\y{T}}$ using the information contained in the complete time series
match remarkably closely those intervals identified in the super-exponential
bubble state with the causal {\itshape filtering probability} $\np{s_{t}}{\y{t}}$ that use data up to the running present time $t$.
This is perhaps not too surprising since our bubble detection method does not rely on the existence of a crash or
change of regime but solely on the identification of a transient super-exponential dynamics. 

$\bullet$ It is quite striking that the periods identified in the super-exponential
bubble state are in general associated with a strong upward price dynamics (not a surprise from the construction of the model)
ending with a peak of the price followed by a clear change of price regime (this second property being much less trivial).
This change can take the form of a significant correction or crash, of a sideway volatile dynamics or of a downward price momentum.
The presence of such changes of regime is a pleasant qualitative confirmation of one of the key
insights of super-exponential models, namely the non-sustainable nature of the price dynamics that has
to break into a slower growth, in fact often a reversal. 

$\bullet$ While most bubble researchers would agree
on the diagnostic of a bubble in a number of cases found to correspond to large values of our smoothing and filtering probabilities,
such as during the price ascent associated with the dot-com bubble in Western markets, 
which is here correctly identified by our method, one could raise the criticism that many periods
are picked out that do not correspond to the times of well-documented bubbles, raising the spectrum
of too many false alarms (false positives or errors of type I). In response to this, we argue that, as said above,
most of these identified bubble periods are followed by a significant change of price regime, which can be
taken as a validation of the bubble identification. Moreover, it is the goal of a better model with superior implementation
to indeed reveal hitherto hidden aspects of the price dynamics. Thus, it should not be a surprise that our method
identifies a number of bubble regimes that would have not been suspected by other means.
Figs. \ref{F:1}-\ref{F:2'} thus lead us to 
conjecture that financial markets are much more often in bubble states that previous believed.

We have quantified the fractions of time, weighted by the corresponding probabilities, that the 16 studied
markets are in a bubble. 
Table\ref{T:fctime} below lists the results, with the fractions expressed as percentages. The notations p.b(filter) and p.b(smoothing) 
indicate the fractions of time when markets are in a bubble regime, as determined respectively by filtering probabilities and smoothing probabilities. 
These fractions are obtained as the integral of the corresponding probabilities shown in Fig. \ref{F:1} - \ref{F:2'} over the whole time period for each of the 16 investigated markets. Except for the Russian market with fractions as low as 8\%, the chance for western stock markets 
to be in bubble regime is typically in the range 10\%-15\% cross-sectionally. 
In contrast, emerging markets as well as Japan market, the only developed market in Asia, are found
about 20\% to 45\% of the time in the bubble regime. This may due to the less 
transparent information environment as well as weaker regulatory control, which tends to promote herding.
 
\begin{table}[ht]
\centering
\caption{Fractions of time that the sixteen studied stock markets are in bubble regime. We give
the fractions with only two digits, as a higher precision would be provide a misleading sense of accuracy.}\label{T:fctime}
\resizebox{8cm}{!}{
\begin{tabular}{c D{.}{.}{2}   D{.}{.}{2}}
  \toprule
 Stock Index & \multicolumn{1}{c}{p.b(filter,\%)} &  \multicolumn{1}{c}{p.b(smoothing,\%)} \\ 
  \midrule
NASDAQ & 15 & 16 \\ 
  SP500 & 12 & 12 \\ 
  FTSE & 10 & 11 \\ 
  RTSI & 8 & 9 \\ 
  CAC40 & 12 & 13 \\ 
  DAX & 11 & 11 \\ 
  MIB & 12 & 13 \\ 
  SMI & 12 & 13 \\ 
  N225 & 40 & 43 \\
  SSEC & 30 & 30 \\ 
  SZSC & 32 & 32 \\ 
  HSI & 24 & 25 \\ 
  STI & 27 & 28 \\ 
  BVSP & 46 & 46 \\ 
  SENSEX & 36 & 38 \\ 
  KS11 & 23 & 23 \\ 
   \bottomrule
\end{tabular}
}
\end{table}

$\bullet$ The previous remark raised the issue of false positives.  What about false negatives or errors of type II?
Some studies have suggested that the run-ups of stock market prices of Western markets from 2003 to 2007 
qualify as a bubble \citep{Sornwood10,SorCauwels2014,SorCauwels2015}. In contrast, Figs. \ref{F:1} and Fig. \ref{F:1'}
give a zero bubble probability for this time interval for all Western markets. One likely explanation of these missed targets
was advanced by \cite{AS2}, who stressed that, by construction the \citep{AS1} model
assumes that bubbles are associated to {\it both} a super-exponential price appreciation {\it and} a corresponding
explosion of the volatility. But \cite{AS2} showed that, for a number of historical bubbles, this is counterfactual:
many bubbles develop without a clear increase of volatility, which would then lead the Sornette-Andersen
model to be rejected. Indeed, many bubbles are developing over a time of complacent view of the underlying risks,
in other words, the genuine risks are under-estimated and the crash often comes as a surprise.
This led  \cite{AS2} to suggest the existence of at least two types of bubbles, the ``fearful'' (resp. ``fearless'') bubbles
associated with an increasing (resp. constant or decreasing) volatility. Our present investigation can thus
be understood as targeting only the fearful bubbles.

\section{Speculative Influence Network}\label{S3}

\subsection{Definitions}

We introduce the {\itshape Speculative Influence Network} (SIN) as a directional weighted network of financial assets, such as stocks, bonds, mutual funds, real estate, forex, commodities, derivatives and so on, which is organized to map the mutual relationships representing the speculative influences between them, i.e., how speculative trading of one asset may draw speculative trading in another assets. The arrow of a link in the SIN then indicates the direction of the speculative trading influence, and its weight quantifies the intensity of the influence. In order to derive the causal speculative influence between assets,  we propose a two-steps procedure: (i) estimate for each asset the daily filtering probability for the presence of super-explosive bubbles
using the HMM based approach developed in Section \ref{S22} with the regime-switching Sornette-Andersen bubble model;
(ii) once two assets are diagnosed as being in a bubble state, identify the possible causal flow of the probabilities from one asset to the
other in order to infer a possible influence or contagion effect. In other words, provided one asset $X$ is diagnosed to be in a bubble, 
we want to estimate how much does it influences the probability of another asset $Y$ to be also in the bubble state.

There are at least two well-known methods to detect causal relationship between time-series, the Granger Causality test and the information-based Transfer Entropy (TE) formulation. The former is relatively ``cheaper'', being linear by construction, while the later encompasses nonlinear
relationship and does not rely on a model specification. Although the Granger Causality test is more popular in economics and finance,  it is 
mainly designed for multivariate linear autoregressive processes with white-noise residuals. For general nonlinear stochastic time series, it 
provides only an approximate measure of causal influence \citep{barrett2013granger}. Moreover, elegant proofs show
that the two methods are only equivalent for Gaussian, exponential Weinman and log-normal variables \citep{barnett2009granger, hlavackova2011equivalence}. Our inputs are the time series of the {\itshape filtering probabilities} $\np{s_{t}}{\y{t}}$ for
all assets, which exhibit properties far from the conditions of application of the Granger Causality test: (i) the $\np{s_{t}}{\y{t}}$'s
are bounded in $[0,1]$; (ii) they exhibit strong non-Gaussian characteristics with a pronounced bimodality near $0$ or $1$;
there is no guarantee that they obey autoregressive processes. Therefore, we prefer to use the TE method in the remaining of the article.

\subsection{Transfer Entropy }\label{S31}

In his famous work ``A Mathematical Theory of Communication", Shannon pioneered a novel metric to quantify the uncertainty of an outcome from a set of possible events, the so-called Shannon Entropy. In order to capture the average uncertainty in a system that is comprised of a set of events with occurrence probabilities $p_i, i=1,\cdots,n$, the Shannon Entropy (SE) $H(p_1,p_2,\cdots,p_n)$ is required to satisfy the following three properties: 
\begin{enumerate}
\item $H(p_1,\cdots,p_n)$ should be continuous in $p_i$.
\item If $p_i=\dfrac{1}{n}, i=1,\cdots,n$,  then $H(p_1,\cdots,p_n)$ should monotonically increase with $n$.
\item If an event $i$ can be further broken down into several sub-events, then the SE is updated by adding the sums of 
the SE of such event with weights equal to their occurrence probabilities, i.e., $H_{\text{new}}=H_{\text{old}}+p_i H_i$, where $H_i$ is the value of the function $H$ for the composite event.
\end{enumerate}

Then, it can be proved that these three properties lead to the unique Shannon Entropy function  $H=\displaystyle -\sum_{i=1}^N p_i\log_s p_i=-\sum_{i=1}^N \p{i}\log_s \p{i}$, where the base $s$ of the logarithm is arbitrary and depends on the chosen information units. For example, $s$ is often selected to be equal to $2$ when the measure is given in terms of bits of information. Shannon Entropy bears a lot of resemblance to Gibbs' entropy, which quantifies the degree of diversity for a system's possible micro-states. But SE is more general and gauges
the necessary external information inflow needed to ascertain a specific micro-state for a event to happen. 

When dealing with systems that interact with each other, the notion of \emph{Transfer Entropy (TE)} can be introduced to describe the extent 
to which the uncertainty of one system is influenced by other systems inter-temporally. Consider two systems $U$ and $V$ that are Markov processes of degree $k_U$ and $k_V$ respectively. Let us denote the sets of possible events for $U$ and $V$ at time $t$ by $u_t$ and $v_t$. Then, the TE from system $V$ to $U$ at time $t$ is defined as 
\begin{align}
\text{TE}_{V\to U}(t)=&-\sum_{\substack{u_t,\cdots,u_{t-k_U}\\ v_{t-1},\cdots,v_{t-k_V}}} \p{u_t,\cdots,u_{t-k_U},\, v_{t-1},\cdots,v_{t-k_V}}\cdot\log _s \np{u_t }{u_{t-1},\cdots,u_{t-k_U}}\notag\\
\phantom{=}&-\left(-\sum_{\substack{u_t,\cdots,u_{t-k_U}\\ v_{t-1},\cdots,v_{t-k_V}}} \p{u_t,\cdots,u_{t-k_U},\, v_{t-1},\cdots,v_{t-k_V}}\cdot\log _s \np{u_t}{u_{t-1},\cdots,u_{t-k_U},v_{t-1},\cdots,v_{t-k_V}}\right)\notag\\
=& \sum_{\substack{u_t,\cdots,u_{t-k_U}\\ v_{t-1},\cdots,v_{t-k_V}}} \p{u_t,\cdots,u_{t-k_U},\, v_{t-1},\cdots,v_{t-k_V}}\cdot\log _s \dfrac{\np{u_t}{u_{t-1},\cdots,u_{t-k_U},v_{t-1},\cdots,v_{t-k_V}}}{\np{u_t }{u_{t-1},\cdots,u_{t-k_U}}}\label{E:tecal}
\end{align}

According to the above definition, the TE from $V$ to $U$ is just the difference between the amount of uncertainty for $U$ when merely measured based on its past information and its amount of uncertainty when also taking account of $V$'s past information. Hence, the TE effectively measures the reduction of uncertainty of system $U$ achieved by using the information of system $V$. In other words, it 
provides an evaluation of the power of $V$ to predict the motion of $U$. It quantifies the strength of the causal relationship 
between $U$ and $V$. In the following subsection, we show how to use the TE to develop the SIN for financial markets.

\subsection{Constructing the {\itshape Speculative Influence Network} (SIN) for a financial market}\label{S32}

In the present work, the speculative influence relationship between two different financial assets is measured by calculating the TE between the time-series of the filtering probabilities defined in subsection \ref{S22} and illustrated in subsection \ref{wrntuyjnhwg},
conditional on the two assets being in the bubble state. For this, we 
introduce the \emph{speculative influence intensity} (SII) from $X$ to $Y$ as 
\begin{equation}
SII_{X\to Y}(t)=TE_{p^f_b(X)\to p^f_b(Y)}(t)~,
\end{equation}
where the time-series $p^f_b(X)=\bigl\{\np{s_t=1}{\ln p_{X,t},\ln p_{X,t-1},\cdots, \ln p_{X,0} }\bigr\}_{t=0}^{t=T}=\bigl\{\np{s_t=1}{x^{t}_{\multimap}}\bigr\}_{t=0}^{t=T}$ and $p^f_b(Y)=\bigl\{\np{s_t=1}{\ln p_{Y,t},\ln p_{Y,t-1},\cdots, \ln p_{Y,0} }\bigr\}_{t=0}^{t=T}=\bigl\{\np{s_t}{y^{t}_{\multimap}}\bigr\}_{t=0}^{t=T}$ respectively denote the estimated filtering probability of staying in the bubble state for assets $X$ and $Y$.

According to the derivations presented in subsection \ref{S22}, the filtering probability $\np{s_{t}}{\y{t}}$ at time $t$ only depends on 
the previous one $\np{s_{t-1}}{\y{t-1}}$ at $t-1$ and on $y_t$. This is intended to capture the idea that the level 
of herding at a given time is mostly influenced by that of the previous time period (day in our empirical estimations). Using this property, 
formula \eqref{E:tecal} for the TE of $V$ to $U$ can be simplified into
\begin{align}
\text{TE}_{V\to U}(t)=&\sum_{\substack{u_t,u_{t-1}, v_{t-1}}} \p{u_t, u_{t-1},v_{t-1}}\cdot 
\log _s \frac{\np{u_t}{u_{t-1},v_{t-1}}}{\np{u_t}{u_{t-1}}} \notag\\
=&\sum_{\substack{u_t,u_{t-1}, v_{t-1}}} \p{u_t, u_{t-1},v_{t-1}}\cdot \log _s \frac{\p{u_t, u_{t-1},v_{t-1}}\p{u_{t-1}}}{\p{u_t, u_{t-1}}\p{u_{t-1},v_{t-1}}}\label{E:tecalsimp}
\end{align}
where $U$ and $V$ actually represent $p^f_b(X)$ and $p^f_b(Y)$, and correspondingly $u_t=\np{s_t=1}{x^{t}_{\multimap}}$ and $v_t=\np{s_t=1}{y^{t}_{\multimap}}$. 

In order to use \eqref{E:tecalsimp}, we need to specify $\p{u_{t-1}}, \p{u_{t-1},v_{t-1}}, \p{u_t,u_{t-1}}$ and $\p{u_t, u_{t-1},v_{t-1}}$.
For this, we divide the range $[0,1]$ of $p^f_b(X)$ and $p^f_b(Y)$ in $10$ bins of equal width and treat each bin as a unique identified event for gauging the presence of speculative herding. This coarse-graining procedure amounts to considering two events within a given bin 
as being undistinguishable and is performed to reduce the sensitivity to noise and minimize the impact of model mis-specification. Correspondingly, the base $s$ in \eqref{E:tecalsimp} is set to 10 for the practical computing.
Then, by counting the number of different combinations associated with the value vector of time-series $U$ up to a certain day,
with the vector for $U$'s values lagged by one day and with the vector for $V$'s values also lagged by one day, 
all four distributions $\p{u_{t-1}}, \p{u_{t-1},v_{t-1}}, \p{u_t,u_{t-1}}$ and $\p{u_t, u_{t-1},v_{t-1}}$ can be empirically calculated.  
$\p{u_{t-1}}$ is obtained by taking the ratio of the number of times a $\p{u_{t-1}}$ in a given bin appeared in the vector $U$ 
by the total number of occurrences for all bins types. In order to estimate $\p{u_{t-1},v_{t-1}}$, one must count how many times a particular combination of bins, that the couple states $(u,v)$ belongs to, appeared in the joint vector of lagged $U$'s values and lagged $V$'s values, and then 
normalise by the total number of occurrences for the pairs. The calculation of $\p{u_t, u_{t-1},v_{t-1}}$ is similar to
that of $\p{u_{t-1},v_{t-1}}$, except that the counting is conducted on bins triplets that appear in the vector 
obtained by joining all three vectors. The calculation of SII between arbitrary pairs of two investment targets finally gives 
the SII matrix for the whole  financial market,  which is the basic encoding scheme to derive the adjacency matrix
that defines the SIN.

The assets of interest are the nodes of the SIN. A relationship in which node $X_i$ influences node $X_j$ is quantified by
$\text{SII}_{X_i\to X_j}$ and is represented by a directed link $X_i \to X_j$. This does not exclude the possible
existence of a direct influence in the opposite direction from node $X_j$ to node $X_i$ measured by $\text{SII}_{X_j\to X_i}$,
making the SIN a bilateral directional network. In practice, we use a threshold so that values of $\text{SII}_{X_i\to X_j}$
below this threshold are considered too small to be of significance and only the transfer entropies larger than this
threshold will be used in the SIN representation. To further extract the leading influence relationship, we 
introduce the {\it net speculative influence intensity} (NSII)
\begin{equation}
\text{NSII}_{\,X_i \to X_j}(t)=\text{SII}_{X_i\to X_j}(t)-\text{SII}_{X_j\to X_i}(t)~,
\label{etyjkikiu}
\end{equation}
allowing us to reduce SIN to an unidirectional network capturing the net influence effects between assets.
In this reduced unidirectional SIN, only positive NSII values need to be considered and are represented 
as arrows going from the influencer $i$ to the receiver $j$ (for $\text{NSII}_{\,X_i \to X_j}(t)>0$).

\section{Early warnings of bubbles for Chinese Markets via SIN}\label{S4}
\subsection{Data and methodology}\label{41}

We now investigate empirically how the Speculative Influence Network (SIN) can help provide 
effective early warning signals by dissecting a bubble structure via 
a disaggregated firm level analysis. We show below that the SIN complements the more
conventional bubble detection method based on super-exponential growth signatures of the 
aggregated market, by providing indicators of strong speculative influence (or contagion)
that can be useful by playing the role of systemic risk gauges. Our tests consist in quantifying the 
out-of-sample performance of the prediction of the maximum cumulative lost proportion (\%{MaxLoss}) 
of each asset based on the SIN analysis.

We focus our attention on the Chinese stock market and on the special periods from 2006 to 2008 when the stock market 
exhibited remarkable signatures of speculation and bubble behavior. \cite{Jiang2010} have provided
a detailed synthesis of this episode, using the log-periodic power law singularity model.
Following a bearish trend that lasted nearly 5 years, the Chinese stock market rebounded in 2005.
The most representative stock index for the market of A-shares, the Shanghai Stock Exchange Composite index (SSEC), 
reached its lowest point in 2005 at the level 998. It rose exuberantly up to 6124 (corresponding to a relative appreciation of 513.5\%) over a mere two years interval. Meanwhile, the second most important index, the Shenzhen Stock Exchange Component index (SZSC), reached a peak of 19600 from its historical lowest point at 3372 (corresponding to a relative appreciation of
481.2\%) over just eighteen months. Correspondingly, the
total market value for all stocks traded  both in the Shanghai Stock
Exchange (SHSE) and in the Shenzhen Stock Exchange (SZSE) has grown explosively 
from 30 trillion yuan at the beginning of 2005 to 
about 250 trillion yuan at the peak in October 2007. This made China's stock market temporarily
the World's third largest market.
Since its peak in 17th October 2007 at 6124, the SSEC has dropped
to the low of 1664 on 28th October 2008. Similarly, after its peak 
at 19600 on the 10th of October 2007, the SZSC reached its bottom 
at 5577.33 on the 28th of October 2008. For both markets, the total drawdown from peak-to-valley
corresponds to a loss in excess of 70\% in just one year, surpassing 
all the losses incurred during the bearish period from 2001 to 2005. 
Such impressive and breathtaking roller-coaster performance with extraordinary 
quintupling of the value of the main Chinese financial indices in less than two years 
accompanied by a rapid shrinkage in just one year back to a longer trend make the Chinese
stock market an ideal specimen on which to tests our proposed analytics.

For the purpose of constructing the early warning signals, the SIN of the Chinese stock market is constructed within the window 
from 2006 to the end of 2007, covering the period when the market is still in its growth regime,
without significant large financial sell-offs. The assets we consider to construct the SIN are the 
Sector Series from the CSI 300 index, which are compiled by the China Securities Index Company, calculated since April 8, 2005 and designed to replicate the performance of 300 stocks traded either in SHSE or in SZSE with outstanding liquidity and size. 
The sector series are sub-indices based on the stocks in the CSI 300 that reflect specific industrial sectors. There are ten CSI 300 sector indices in total: Energy, Materials, Financial, Industrials, Consumer Discretionary, Consumer Staples, Health Care, Information Technology, Telecommunications and Utilities. Because of the special role played by 
the financial sector in terms of its almost unique large leverage, its essential role as credit provider to the economy, and given 
its recognized responsibility in propagating systemic shocks to other sectors of the economy, we disaggregate 
the CSI 300 Financial Index into its individual firm components, according to the classification by China Securities Regulatory Commission. 
 
Table \ref{T:inddesc} lists some basic descriptive statistics for the nine industrial indices covered in this article. In the putative bubble phase starting from 2006 to the end of 2007, the table gives the maximum and minimum values of those indices as well as their daily return $\mu$ and volatility $\sigma$ (estimated based on a specification in terms of a geometric Brownian motion and shown in percentage form). The sixth and seventh columns respectively show the fraction of filtering probabilities calculated daily (based
on the method presented in subsections \ref{S22} and \ref{wrntuyjnhwg})
that meet particular criteria over the period. Specifically, column $\text{h.f.p(\%)}$ gives the percentage of times when the filtering probability is larger than $0.9$. Column $\text{l.f.p(\%)}$ gives the percentage of times when the filtering probability is 
smaller than $0.1$, indicating that the speculative bubble behavior is fading away on that sector while it is active
for the overall market. The last column (\%{MaxLoss}) gives the maximum percentage loss for 
each industrial sector over 2008. This loss is defined as the RMB amount of the maximum cumulative decline in market capitalization of the corresponding representative stock index during 2008 divided by the maximum market capitalization in 2008. 
Similarly, Table \ref{T:findesc} lists the same basic descriptive statistics for the financial firms constituting the Financial CSI 300 sector index. There are 22 financial stocks respectively belonging to four sub-sectors:  bank, securities, trust and insurance. 
Comparing the two tables, it is quite impressive to note the much larger variations of these basic statistics 
at the disaggregated financial firm level in Table \ref{T:findesc}  compared to the non-financial sectors level in 
Table \ref{T:inddesc}. One can also observe that the stocks of securities or trust companies
exhibit larger daily volatility than bank and insurance companies. However, such facts have little 
relation with the observed proportion of speculative days $\text{h.f.p(\%)}$ and with \%{MaxLoss}.         

\begin{table}[ht]
\centering
\caption{Descriptive statistics for the nine chosen representative Chinese industrial indices.
Column $\text{h.f.p(\%)}$ gives the percentage of times when the filtering probability is larger 
than $0.9$. Column $\text{l.f.p(\%)}$ gives the percentage of times when the filtering probability is 
smaller than $0.1$. The last column (\%{MaxLoss}) gives the maximum percentage loss for 
each industrial sector over 2008 as defined in the text.}\label{T:inddesc}
\resizebox{15cm}{!}{
\begin{tabular}{lccccccc}
  \toprule
\multirow{3}*[2.5mm]{Industrial Sectors}&\multicolumn{6}{c}{time window: 2006-2007}& 2008\\
\cline{2-7}
 & Min & Max & $\mu$(\%)$\phantom{\displaystyle\sum}$ & $\sigma$(\%) & h.f.p.(\%) & l.f.p.(\%) & \%{MaxLoss} \\ 
  \midrule
Energy & 1078.1 & 7740.5 & 0.37 & 2.32 & 40.90 & 48.47 & 73.75 \\ 
  Materials & 868.3 & 6419.4 & 0.38 & 2.23 & 27.64 & 59.23 & 76.03 \\ 
  Industrials & 884.0 & 5342.9 & 0.36 & 2.04 & 26.60 & 61.34 & 73.35 \\ 
  Consumer Discretionary & 893.4 & 5332.9 & 0.35 & 2.19 & 29.62 & 59.10 & 71.95 \\ 
  Consumer Staples & 1052.1 & 7501.9 & 0.41 & 2.06 & 28.71 & 59.66 & 65.91 \\ 
  Health Care & 833.7 & 5242.2 & 0.37 & 2.28 & 28.07 & 62.38 & 59.65 \\ 
  Information Technology & 727.5 & 2694.1 & 0.26 & 2.27 & 28.80 & 54.46 & 74.31 \\ 
  Telecommunications & 950.6 & 4163.0 & 0.29 & 2.36 & 20.54 & 72.39 & 67.29 \\ 
  Utilities & 879.5 & 3738.9 & 0.28 & 2.19 & 30.31 & 60.01 & 60.07 \\ 
   \bottomrule
\end{tabular}
}
\end{table}

\begin{table}[ht]
\centering
\caption{Descriptive statistics for 22 chosen stock prices belonging to the financial sector. Same definitions 
as Table \ref{T:inddesc}.}\label{T:findesc}
\resizebox{18cm}{!}{
\begin{tabular}{llcccccccc}
  \toprule
\multirow{3}*[2.5mm]{Financial Institution}&\multirow{3}*[2.5mm]{Abbr.}&\multirow{3}*[2.5mm]{Sub-sectors}&\multicolumn{6}{c}{window: 2006-2007}& 2008\\
\cline{4-9}
 & & & Min & Max & $\mu$(\%)$\phantom{\displaystyle\sum}$ & $\sigma$(\%) & h.f.p.(\%) & l.f.p.(\%) & \%{MaxLoss} \\ 
  \midrule
  Ping An Bank & PAB & Bank & 165.9 & 1440.2 & 0.44 & 3.26 & 27.10 & 67.32 & 75.43 \\ 
  Hong Yuan Securities & HYS & Securities & 12.0 & 202.2 & 0.59 & 4.25 & 20.75 & 71.48 & 76.23 \\ 
  Shaanxi International Trust & SIT & Trust & 11.4 & 111.8 & 0.45 & 4.21 & 11.81 & 82.38 & 73.51 \\ 
  Northeast Securities & NES & Securities & 9.0 & 369.2 & 1.26 & 12.38 & 8.75 & 85.41 & 81.61 \\ 
  Guoyuan Securities Company & GSC & Securities & 2.9 & 61.8 & 1.26 & 10.30 & 11.42 & 81.66 & 78.27 \\ 
  Changjiang Securities Company & CJSC & Securities & 4.9 & 76.9 & 1.18 & 10.63 & 30.15 & 60.79 & 77.17 \\ 
  Shanghai Pudong Development Bank & SPDB & Bank & 16.0 & 125.1 & 0.44 & 3.20 & 34.75 & 55.88 & 76.66 \\ 
  Hua Xia Bank & HXB & Bank & 6.2 & 40.2 & 0.40 & 3.34 & 23.39 & 67.71 & 71.06 \\ 
  China Minsheng Banking Corp. & CMBC & Bank & 19.3 & 146.2 & 0.37 & 2.89 & 26.38 & 64.19 & 68.69 \\ 
  Citic Securities Company & CSC & Securities & 7.2 & 162.9 & 0.62 & 3.79 & 14.87 & 80.84 & 66.84 \\ 
  China Merchants Bank & CMB & Bank & 13.4 & 110.0 & 0.43 & 2.78 & 28.63 & 60.66 & 73.86 \\ 
  Sinolink Securities & SLS & Securities & 4.5 & 166.7 & 0.88 & 5.92 & 13.68 & 82.47 & 69.85 \\ 
  Southwest Securities & SWS & Securities & 2.1 & 18.3 & 0.96 & 3.20 & 24.15 & 71.95 & 78.45 \\ 
  Anxin Trust & AXT & Trust & 12.9 & 152.2 & 0.78 & 4.75 & 11.52 & 84.59 & 72.66 \\ 
  Haitong Securities Company & HSC & Securities & 38.0 & 626.8 & 0.61 & 4.33 & 6.99 & 89.92 & 75.51 \\ 
  Industrial Bank & CIB & Bank & 22.2 & 67.7 & 0.40 & 3.24 & 26.78 & 66.20 & 80.68 \\ 
  Ping An Insurance (Group) & PAIC & Insurance & 44.4 & 145.7 & 0.40 & 3.04 & 38.04 & 54.41 & 80.77 \\ 
  Bank Of Communications & CCB & Bank & 10.4 & 16.8 & 0.09 & 2.78 & 21.62 & 72.99 & 73.09 \\ 
  Industrial And Commercial Bank & ICBC & Insurance & 3.3 & 8.9 & 0.32 & 2.74 & 22.88 & 71.18 & 57.76 \\ 
  China Life Insurance Company & CLIC & Insurance & 32.0 & 75.3 & 0.17 & 3.24 & 29.33 & 61.64 & 68.72 \\ 
  Bank Of China & BOC & Bank & 3.2 & 7.5 & 0.15 & 2.41 & 24.59 & 65.31 & 56.69 \\ 
  China Citic Bank Corp. & CITIC & Bank & 8.5 & 12.6 & -0.07 & 2.97 & 30.83 & 59.90 & 64.21 \\ 
   \bottomrule
\end{tabular}
}
\end{table}

In order to exploit the SIN analysis, we define a set of indicators that measure 
the speculative influence effects at diverse levels of aggregation.
The set made of the nine industrial sector nodes is denoted by $\mathcal{G}_I$.
The set made of the 22 financial institution nodes is denoted by $\mathcal{G}_F$.
\begin{enumerate}
\item \emph{SI-to-All} measures the total speculative influence of a given node $i$ to all
other nodes, including the industrial sectors and financial institutions:
\begin{equation}
\text{SI-to-All}_i=\sum_{\substack{j\in \mathcal{G}_I\cup\mathcal{G}_F\\ j\ne i}} \text{SII}_{i\to j}~.
\label{SI-ind1}
\end{equation}
This indicator will allow us to identify those nodes that have a significant global influence on the entire network.

\item \emph{SI-from-All} measures the total speculative influence of all other nodes including industrial sectors and financial institutions on a given node $i$:
\begin{equation}
\text{SI-from-All}_i=\sum_{\substack{j\in \mathcal{G}_I\cup\mathcal{G}_F\\ j\ne i}} \text{SII}_{j\to i}~.
\label{SI-ind2}
\end{equation}
This indicator identifies the sectors or firms that are the most susceptible to the overall speculative influence of the network.

\item \emph{SI-to-Fin} measures the total speculative influence of a given node $i$
to all financial institution nodes:
\begin{equation}
\text{SI-to-Fin}_i=\sum_{\substack{j\in \mathcal{G}_F\\ j\ne i}} \text{SII}_{i\to j}~.
\label{SI-ind3}
\end{equation}
This indicator quantifies how certain institution and sector's speculative behavior may influence the financial sector firms.

\item \emph{SI-from-Fin}  measures the total speculative influence of all financial nodes onto a given node $i$:
\begin{equation}
\text{SI-from-Fin}_i=\sum_{\substack{j\in \mathcal{G}_F\\ j\ne i}} \text{SII}_{j\to i}~.
\label{SI-ind4}
\end{equation}

\item \emph{SI-to-IX} measures the total speculative influence of a given node $i$ to all
other nodes of the industrial sectors (excluding the financial institutions):
\begin{equation}
\text{SI-to-IX}_i=\sum_{\substack{j\in \mathcal{G}_I\\ j\ne i}} \text{SII}_{i\to j}~.
\label{SI-ind5}
\end{equation}

\item \emph{SI-from-IX} measures the total speculative influence of all other nodes of the industrial sectors (excluding
the financial sector) on a given node $i$:\begin{equation}
\text{SI-from-IX}_i=\sum_{\substack{j\in \mathcal{G}_I\\ j\ne i}} \text{SII}_{j\to i}~.
\label{SI-ind6}
\end{equation}
\end{enumerate}

In addition, using the {\it net speculative influence intensity} (NSII) defined by (\ref{etyjkikiu}) leading to 
a unidirectional network, we shall also use the following indicators obtained from (\ref{SI-ind1}-\ref{SI-ind6})
as follows:
\begin{align}
\text{NSII-on-All}_i=&\sum_{\substack{j\in \mathcal{G}_I\cup\mathcal{G}_F\\
 j\ne i}}\text{NSII}_{i \to j}=\text{SI-to-All}_i-\text{SI-from-All}_i   \\
\text{NSII-on-Fin}_i=&\sum_{\substack{j\in \mathcal{G}_F\\ j\ne i}} \text{NSII}_{i \to j}=\text{SI-to-Fin}_i-\text{SI-from-Fin}_i\\
\text{NSII-on-IX}_i=&\sum_{\substack{j\in \mathcal{G}_I\\ j\ne i}} \text{NSII}_{i \to j}=\text{SI-to-IX}_i-\text{SI-from-IX}_i~.
\end{align}
These indicators quantify the net speculative influence flux of a node to all the other nodes ($\text{NSII-on-All}_i$),
to all financial nodes ($\text{NSII-on-Fin}_i$) and to all non-financial nodes ($\text{NSII-on-IX}_i$).

We want to detect what combination of indicators for individual institution/sector performs best as 
the leading factor during a bubble phase to explain the \%{MaxLoss} of a given node associated with a
large financial sell-offs during a bubble collapse. The corresponding leading factors are good
candidates to provide early-warning signals encoded in the SIN and to become systemic risk metrics.

\subsection{Early-warning signal extraction from the speculative influence network (SIN) \label{42}}

To evaluate the predictive power of SIN indicators defined above, we first perform multivariate regressions of \%{MaxLoss} on different combinations of the 6 indicators  (\ref{SI-ind1}-\ref{SI-ind6}). For the purpose of completeness and to avoid unnecessary multicollinearity, 17 regression models in total are selected. On the basis that industrial sectors and financial institutions 
play different roles, we simplify the treatment by performing tests
on group $\mathcal{G}_F$ and group $\mathcal{G}_I$ separately. The results are shown in tables \ref{T:ewse1}  
and \ref{T:ewse2}.   All indicators are constructed with data from 1st January 2006 to 31st December 2007 and they are ranked before implementing the regressions. Without loss of information and to avoid inflating the tables, we do not report the 
intercepts of the regressions.

For Chinese financial institutions, we find that SI-to-Fin and SI-from-Fin are both significant determinants to explain their maximum drawdown in 2008, while their impacts are of opposite signs. The positive sign obtained for the ranked value of SI-to-Fin means that the more powerful is the speculative herding influence of a node towards a financial institution, the larger is the 
loss of that node during the bubble burst. These results suggest the institutions with higher SI-to-Fin rank are more likely to act as the leading speculative engines of the overall speculative mania within the financial sector, with more inflow of speculative money during bubble. The negative sign obtained for the ranked value of SI-from-Fin 
implies that the larger is the speculative influence of financial institutions towards a given node, the stronger
is the resilience of that node during the financial sell-offs. 
The institutions with higher SI-from-Fin rank are more likely to be
lagging in the speculative frenzy and may be less exposed to the bubble collapse. 

Table \ref{T:ewse1} demonstrates that SI-to-IX has also a high explanatory power for the loss sizes 
of industrial institutions (excluding financial firms). However, different from SI-to-Fin, the larger the 
speculative influence of a node on other industrial sector firms, the smaller is the loss during the crash.
In contrast, SI-from-IX has an insignificant effect, suggesting that there is little transmission of 
speculation from the industry sectors to other firms.
Additionally, table \ref{T:ewse1} also shows that SI-to-IX has a high explanatory power for the loss sizes 
of financial institutions. Again, different from SI-to-Fin, the larger the speculative influence of a node on all  nodes of the industrial sectors, the smaller is the loss during the crash for this financial institution. We also observe that
the risk transfer between financial institutions and industrial sectors is not symmetric. Table \ref{T:ewse2} shows
that there is little influence of the variable SI-to-Fin onto the industrial sector concerning its loss size. 
Meanwhile, given the insignificance of the variable SI-from-IX to explain the losses of financial institutions shown in table \ref{T:ewse1}, we
conjecture that, during the bubble, the Chinese financial sector as a whole only provided speculative money to other industrial sectors 
but did not absorb it back during the crisis.

As shown in table \ref{T:ewse2}, SI-to-IX and SI-from-IX both become significant determinants 
of the losses of the industrial sector in 2008. This is very similar to the results 
obtained for financial institutions listed in table \ref{T:ewse1}. This also suggests that the nodes
that have the leading speculative influence on their peers bear the largest loss risks, while
the followers are less punished by the market correction.
However, we failed to find SI-from-Fin to be a significant indicator for \%{MaxLoss} of industrial sectors. At first glance, 
this appears to be in contradiction with our previous conjecture that speculative money is more likely to go to industrial sectors, 
and taken out of financial institutions that have more influence onto industrial sectors as a whole. 
Noting that the financial sector as a whole is nothing but a peer node within the network of all industrial sectors,
it is possible that the total flow of speculative money emanating from the financial sector is not 
sufficient to increase the risk exposure of the industrial sectors for this Chinese bubble.
The \%{MaxLoss} of industrial sectors should be in large part explained by 
the overall circulation of speculative money among all the nodes within the network rather than by 
the flow to and from a single node, such as the financial sector. In other words, this shows the limit
of a simple causal analysis of influences, suggesting the importance of a complex nonlinear set of
mutual interactions that cannot be simply linearly disentangled. In hindsight, this 
interpretation is also partially supported by the lack of evidence of the
presence of systemic risks spreading from the financial sector to other Chinese economic sectors
through illiquidity, insolvency, or losses during the bubble collapse.  

\begin{sidewaystable}
\begin{center}
\caption{\small Test of the explanatory power of combinations of indicators described in the text from 1st Jan. 2006 to 31st Dec. 2007 to predict its Max\% Loss over 2008 of the set of 22 financial institution nodes $\mathcal{G}_F$, with a multiple regression analysis. There are six candidate explanatory variables to be chosen: SI-from-All, SI-to-Fin,SI-from-Fin, SI-to-IX,SI-from-IX. In practice, these explanatory variables are 
transformed into their cross-sectionally ranked value before the regressions.} \label{T:ewse1}
\resizebox{!}{3.5cm}{
\begin{tabular}{l c c c c c c c c c c c c c c c c c }
\toprule
            & Model1 & Model2 & Model3 & Model4 & Model5 & Model6 & Model7 & Model8 & Model9 & Model10 & Model11 & Model12 & Model13 & Model14 & Model15 & Model16 & Model17 \\
\midrule
SI-to-All   & $-0.25$  &          &          &                       &          &                      &                       &          &          &                       &                      &          &                        &          &          &          &                       \\
            & $(0.23)$ &          &          &                       &          &                      &                       &          &          &                       &                      &          &                        &          &          &          &                       \\
SI-from-All &          & $-0.29$  &          &                       &          &                      &                       &          &          &                       &                      &          &                        &          &          &          &                       \\
            &          & $(0.22)$ &          &                       &          &                      &                       &          &          &                       &                      &          &                        &          &          &          &                       \\
SI-to-Fin   &          &          & $-0.18$  &                       &          &                      & $0.03$                & $0.17$   & $-0.11$  &                       &                      &          & $\mathbf{1.06}^{*}$    & $0.02$   & $0.25$   &          & $\mathbf{1.17}^{*}$   \\
            &          &          & $(0.23)$ &                       &          &                      & $(0.22)$              & $(0.55)$ & $(0.22)$ &                       &                      &          & $(0.52)$               & $(0.23)$ & $(0.53)$ &          & $(0.58)$              \\
SI-from-Fin &          &          &          &                       & $-0.23$  &                      &                       & $-0.38$  &          & $-0.13$               &                      & $-0.17$  & $\mathbf{-1.05}^{**}$  &          & $-0.39$  & $-0.13$  & $\mathbf{-1.15}^{**}$ \\
            &          &          &          &                       & $(0.23)$ &                      &                       & $(0.55)$ &          & $(0.21)$              &                      & $(0.22)$ & $(0.49)$               &          & $(0.53)$ & $(0.21)$ & $(0.54)$              \\
SI-to-IX    &          &          &          & $\mathbf{-0.53}^{**}$ &          &                      & $\mathbf{-0.54}^{**}$ &          &          & $\mathbf{-0.50}^{**}$ & $\mathbf{-0.48}^{*}$ &          & $\mathbf{-0.74}^{***}$ & $-0.49$  &          & $-0.46$  & $\mathbf{-0.86}^{**}$ \\
            &          &          &          & $(0.20)$              &          &                      & $(0.22)$              &          &          & $(0.21)$              & $(0.27)$             &          & $(0.23)$               & $(0.30)$ &          & $(0.28)$ & $(0.32)$              \\
SI-from-IX  &          &          &          &                       &          & $-0.39$ &                       &          & $-0.37$  &                       & $-0.07$              & $-0.36$  &                        & $-0.07$  & $-0.37$  & $-0.07$  & $0.14$                \\
            &          &          &          &                       &          & $(0.22)$             &                       &          & $(0.22)$ &                       & $(0.27)$             & $(0.22)$ &                        & $(0.28)$ & $(0.23)$ & $(0.28)$ & $(0.28)$              \\
\midrule
R$^2$       & 0.06     & 0.08     & 0.03     & 0.26                  & 0.05     & 0.14                 & 0.26                  & 0.05     & 0.15     & 0.27                  & 0.26                 & 0.17     & 0.41                   & 0.26     & 0.18     & 0.27     & 0.42                  \\
Adj. R$^2$  & 0.01     & 0.03     & -0.02    & 0.22                  & -0.00    & 0.10                 & 0.18                  & -0.05    & 0.06     & 0.20                  & 0.18                 & 0.08     & 0.31                   & 0.14     & 0.04     & 0.15     & 0.28                  \\
F statistic & 1.24     & 1.64     & 0.58     & 6.90                  & 0.99     & 3.24                 & 3.29                  & 0.52     & 1.68     & 3.56                  & 3.33                 & 1.88     & 4.11                   & 2.11     & 1.27     & 2.27     & 3.02                  \\
\bottomrule
\multicolumn{18}{l}{\small{$^{***}p<0.01$, $^{**}p<0.05$, $^*p<0.1$}}
\end{tabular}

}

\bigskip\bigskip
\caption{\small Test of the explanatory power of combinations of indicators described in the text from 1st Jan. 2006 to 31st Dec. 2007 to predict its Max\% Loss over 2008 of the set of nine industrial sector nodes $\mathcal{G}_I$, with a multiple regression analysis. There are six candidate explanatory variables to be chosen: SI-from-All,SI-to-IX,SI-from-IX,SI-to-Fin,SI-from-Fin. In practice, these explanatory variables are transformed into their cross-sectionally ranked value before the regressions.} \label{T:ewse2}
\resizebox{!}{3.5cm}{
\begin{tabular}{l c c c c c c c c c c c c c c c c c }
\toprule
            & Model1 & Model2 & Model3 & Model4 & Model5 & Model6 & Model7 & Model8 & Model9 & Model10 & Model11 & Model12 & Model13 & Model14 & Model15 & Model16 & Model17 \\
\midrule
SI-to-All   & $0.64$   &          &          &          &          &          &          &          &          &          &                      &          &          &                      &          &                     &                     \\
            & $(0.82)$ &          &          &          &          &          &          &          &          &          &                      &          &          &                      &          &                     &                     \\
SI-from-All &          & $0.57$   &          &          &          &          &          &          &          &          &                      &          &          &                      &          &                     &                     \\
            &          & $(0.83)$ &          &          &          &          &          &          &          &          &                      &          &          &                      &          &                     &                     \\
SI-to-IX    &          &          &          & $1.09$   &          &          & $1.11$   &          &          & $1.05$   & $\mathbf{2.48}^{**}$ &          & $1.20$   & $\mathbf{2.48}^{**}$ &          & $\mathbf{2.43}^{*}$ & $\mathbf{2.55}^{*}$ \\
            &          &          &          & $(0.75)$ &          &          & $(0.78)$ &          &          & $(0.81)$ & $(0.91)$             &          & $(0.89)$ & $(0.94)$             &          & $(0.98)$            & $(1.06)$            \\
SI-from-IX  &          &          &          &          &          & $-0.07$  &          &          & $-0.05$  &          & $\mathbf{-1.89}^{*}$ & $-0.12$  &          & $-1.86$              & $-0.08$  & $-1.88$             & $-1.85$             \\
            &          &          &          &          &          & $(0.86)$ &          &          & $(0.90)$ &          & $(0.91)$             & $(0.91)$ &          & $(0.94)$             & $(1.02)$ & $(0.98)$            & $(1.04)$            \\
SI-to-Fin   &          &          & $0.49$   &          &          &          & $0.53$   & $0.33$   & $0.49$   &          &                      &          & $0.96$   & $0.48$               & $0.30$   &                     & $0.84$              \\
            &          &          & $(0.84)$ &          &          &          & $(0.78)$ & $(1.63)$ & $(0.90)$ &          &                      &          & $(1.60)$ & $(0.64)$             & $(1.84)$ &                     & $(1.34)$            \\
SI-from-Fin &          &          &          &          & $0.46$   &          &          & $0.19$   &          & $0.31$   &                      & $0.48$   & $-0.51$  &                      & $0.22$   & $0.29$              & $-0.43$             \\
            &          &          &          &          & $(0.84)$ &          &          & $(1.63)$ &          & $(0.81)$ &                      & $(0.91)$ & $(1.62)$ &                      & $(1.85)$ & $(0.67)$            & $(1.35)$            \\
\midrule
R$^2$       & 0.08     & 0.06     & 0.05     & 0.23     & 0.04     & 0.00     & 0.29     & 0.05     & 0.05     & 0.25     & 0.56                 & 0.04     & 0.30     & 0.60                 & 0.05     & 0.57                & 0.61                \\
Adj. R$^2$  & -0.05    & -0.07    & -0.09    & 0.12     & -0.09    & -0.14    & 0.05     & -0.27    & -0.27    & 0.00     & 0.41                 & -0.27    & -0.12    & 0.36                 & -0.52    & 0.31                & 0.22                \\
F statistic & 0.62     & 0.48     & 0.34     & 2.13     & 0.31     & 0.01     & 1.21     & 0.15     & 0.15     & 1.01     & 3.75                 & 0.14     & 0.72     & 2.50                 & 0.09     & 2.22                & 1.57                \\
\bottomrule
\multicolumn{18}{l}{\small{$^{***}p<0.01$, $^{**}p<0.05$, $^*p<0.1$}}
\end{tabular}}
\end{center}
\end{sidewaystable}

We also performed correlation tests between  \%{MaxLoss} and possible combinations of our chosen indicators. To keep in line with the previously revealed relationships, the test first focuses on NSII and then associates NSII with a possible global influence 
effect for specific sub-networks, without making an exhaustive enumeration of all possible combinations. Similar to the above regression analysis, the tests are implemented separately on group $\mathcal{G}_F$ and group $\mathcal{G}_I$. 
All combinations are ranked before testing. Table \ref{T:coly} presents the results based on three popular correlation statistics. 
We find that NSII-on-IX outperforms other combinations to predict the rank of loss sizes of industrial sectors. By taking account of other overall influence effects, the predictive power decreases in agreement with
 our previous discussion of the regression analysis that the financial sector had
very little influence in transferring risks to other industrial sectors during this bubble. Nevertheless, NSII-on-Fin has
only a modest correlation with the \%{MaxLoss} for financial institutions. However, when considering SI-from-IX, its predictive power increases nearly twofold to become a significant signal. In agreement with our previous inference
that financial institutions do not show significant tendencies to draw ``hot money'' from the real economy sector
but act more as speculative followers, the present correlation estimations confirm that financial institutions
are less important in China with respect to the systemic risks building up during the bubble, which is driven
more by the real economy than the reverse.

\begin{table}[ht]
\centering
\caption{Correlations between \%{MaxLoss} and different combinations of indicators given in the first column, as described in the text.}\label{T:coly}
\begin{tabular}{lrrr}
  \toprule
\multirow{3}*[2.5mm]{Indicators}&\multicolumn{3}{c}{Correlation Statistics}  \\
\cline{2-4} 
& Pearson & Spearman & Kendall\\
  \midrule
  \multicolumn{4}{c}{For Industrial Sectors}\\
  \midrule
  NSII-on-All & -0.06 & -0.07 & -0.06 \\ 
  NSII-on-IX & $\mathbf{0.41}$ & $\mathbf{0.42}$ & $\mathbf{0.28}$ \\ 
  NSII-on-Fin & -0.25 & -0.28 & -0.22 \\ 
  NSII-on-IX$-$(SI-from-Fin) & -0.14 & -0.10 & -0.06 \\ 
  NSII-on-Fin$-$(SI-from-IX) & -0.37 & -0.32 & -0.22 \\ 
  NSII-on-IX$+$(SI-to-Fin) & 0.22 & 0.27 & 0.17 \\ 
  NSII-on-Fin$+$(SI-to-IX) & 0.12 & 0.10 & 0.06 \\ 
  \midrule
  \multicolumn{4}{c}{For Financial Institutions }\\
  \midrule
  NSII-on-All & 0.29 & 0.20 & 0.13 \\ 
  NSII-on-Fin & 0.28 & 0.22 & 0.15 \\ 
  NSII-on-IX & 0.00 & -0.06 & -0.06 \\ 
  NSII-on-Fin$-$(SI-from-IX) & $\mathbf{0.47}$ & $\mathbf{0.39}$ & $\mathbf{0.30}$ \\ 
  NSII-on-IX$-$(SI-from-Fin )& 0.21 & 0.06 & 0.06 \\ 
  NSII-on-Fin$+$(SI-to-IX) & 0.03 & -0.02 & -0.01 \\ 
  NSII-on-IX$+$(SI-to-Fin) & -0.22 & -0.15 & -0.14 \\ 
   \bottomrule
\end{tabular}
\end{table}

\begin{figure}[hbt]
\centering
  \includegraphics[totalheight=15cm]{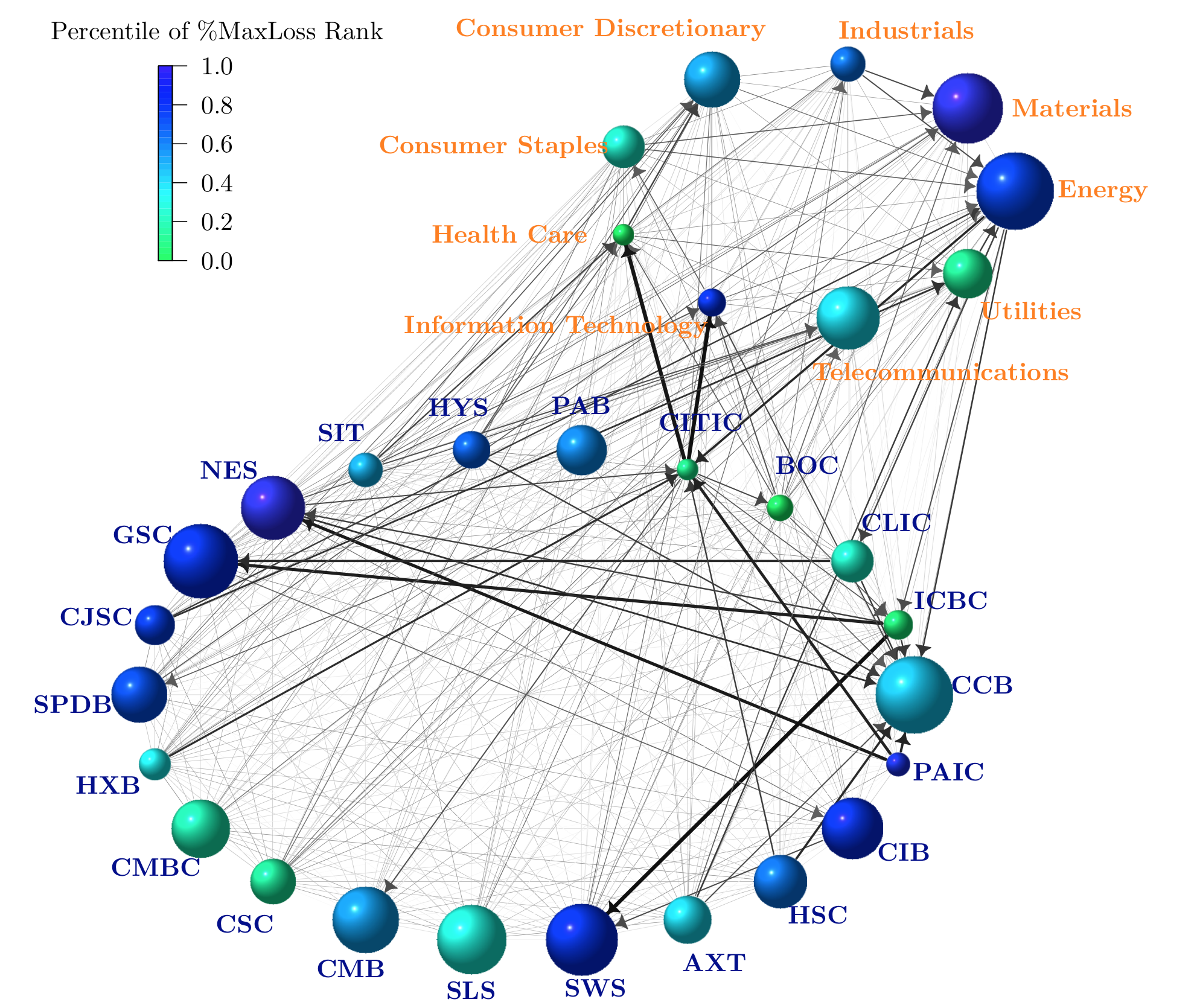}
  \caption{The unidirectional Speculative Influence Network (SIN)  for the Chinese stock market constructed 
  for the time interval from 1st Jan. 2006 to 31st Dec. 2007. This network is comprised of two sub-networks, the industrial sectors with 9 nodes and the financial sub-network with 22 nodes. The arrows connecting nodes represent the speculative influence relationship according to the NSII indicator defined by expression (\ref{etyjkikiu}). The size of a node is determined by its net transfer entropy (TE) from that node to all other nodes.
The color of a node encodes the rank of its \%{MaxLoss} as shown on the scale in the figure inset.}
\label{F:SINex}
\end{figure}

To fully appreciate the impact of the relationships of speculation influence among various industrial sectors as well as financial institutions, Fig. \ref{F:SINex} provides a visualization of the unidirectional SIN calculated by using data in the same time window as before (1st Jan. 2006 to 31 Dec. 2007). This network is comprised of two sub-networks, the industrial sectors with 9 nodes and the financial sub-network with 22 nodes. Recall that this financial network can be regarded as a special node in the former network but
has been disaggregated as explained above. 

First, the size of a node encodes the \emph{net} 
transfer entropy (TE) from that node to all other nodes.  To make the network representation 
comparable with the regression and correlation analyses presented above, we use a slightly different rule
for the sizes of the nodes corresponding to the industrial sectors and to the financial firms. For the industrial sector nodes, 
their sizes are determined by the rank value of their NSII-on-IX.
For the financial institution nodes, their sizes are proportional to the ranked value of their NSII-on-Fin minus SI-from-IX. 

In Fig. \ref{F:SINex}, all nodes are connected by their speculative influence relationship according to the NSII indicator defined by expression (\ref{etyjkikiu}), i.e., an arrow from node $X$ to $Y$ indicates that $X$ has a net positive influence to foster speculative trading on $Y$. The thickness of each arrow encodes the magnitude of the total of 465 NSIIs between 32 nodes. In details, the NSIIs 
are rescaled into $[0,1]$ by a dilation transformation. Then, the  thicknesses of the edges in the SIN are proportional to the rescaled NSIIs. 
To avoid overloading the figure, only the links with NSIIs that are above a threshold are represented with an arrow, as they 
represent the most important speculative influence relationships.
In practice, the threshold was chosen to be  NSII $\geq 0.3$, which yields 58 instances that account for 
approximately 10\% of all links.   

Finally, the color of a node encodes the rank of its \%{MaxLoss} as shown on the scale in the figure inset. 
The two different kinds of nodes (industry and finance) are considered separately when calibrating the mapping function for node size and color.

Based on the colors and sizes of the nodes,  the main information that emerges from Fig. \ref{F:SINex} is
that the sectors or firms that lost the most are those influencing others most. 
The second message of Fig. \ref{F:SINex} is conveyed by the directions and thicknesses of the arrows. 
One can observe that some of the large nodes (e. g. Materials, Energy, CCB) receive many arrows pointing to them.
Given that their sizes encode the fact that they are the largest net influencers, the arrows pointing to them
reveals a second order effect that they also receive large influences (above the threshold NSII $\geq 0.3$) from a few other sectors or firms.
But these large influences only contribute a maximum of 10\% of the total influence relationships. 
Thus, it is more the overall influence over most of the other sectors and firms that dominate the amplitude
of the losses, rather than the (large) influence of a few. Rather than identifying a few systemically important sectors, the SIN analysis
emphasises the role of aggregated speculative influences in determining global interconnectedness and systemic risks.
This interpretation is consistent with the results of the above regression and correlation analyses,
which have concluded that larger Net Transfer Entropies lead to larger losses.    

The case of CITIC is special, as it seems to act as a hub for important two-way speculative influences between financial and industrial sub-networks. 
With smaller NSII-on-Fin minus SI-from-IX effect, this financial institution is, as expected, mildly punished during market correction.  
At first glance, The Information Technology sector seems to be an exception with little net entropy transfer to and from other industrial sectors but nevertheless a relative large loss. This can
however be rationalized by the  important influence relationship from 
the financial sector (in particular CITIC) that might also signal an
invisible ``hot money'' transfer between the sector before the full force of speculative herding 
revealed itself. 
  
\section{Conclusion}\label{S5}

We have introduced the Speculative Influence Network (SIN) to represent the
causal relationships  between sectors (and/or firms) during financial bubbles. 
The construction of SIN required two steps. First, we have shown how to 
estimate in real time the probability of speculative trading in each stock
in the basket of interest, using a bubble identifying technique based on the Sornette-Andersen (2002)
stochastic bubble model. We have augmented the Sornette-Andersen (2002) model 
by using a Hidden Markov Model (HMM),
which was specially designed to calibrate regime-switching between normal and bubble regimes.
Conditional on two stocks or two sectors being qualified in a bubble regime, we have introduced
the {\itshape Transfer Entropy\/}, which quantifies the casual relationship between them.
The transfer entropy provided a natural way to construct an adjacency matrix and thus
to introduce the SIN. 
We have applied this methodology to the Chinese stock market during the period 2005-2008, during which a normal phase
was followed by a spectacular bubble ending in a massive correction. 
Introducing the Net Speculative Influence Intensity (NSII) variable as the difference between the transfer entropies 
from $i$ to $j$ and from $j$ to $i$, we have demonstrated its skill to predict the 
maximum loss (\%{MaxLoss}) endured during the crash by the sectors and stocks.
The statistically significant regressions and the graphical network representation has allowed us to llustrate
the matrix of influences among sectors during the Chinese bubble. One the most striking result 
is the importance of the industrial sectors in influencing the financial firms and not vice-versa,
showing a very different interaction structure than in western markets. We 
have also shown that the bubble state variable
calibrated on the Chinese market data corresponds well to the regimes when the market exhibits a strong
price acceleration followed by clear change of price regimes.

\clearpage

\input{appendix}

\clearpage
\bibliographystyle{elsarticle-harv}
\bibliography{CSSB_HMM_SysRisk}
\end{document}

%% file: preamble.tex
\usepackage{geometry}  
\usepackage{graphicx}
\usepackage{amssymb}
\usepackage{amsmath}
\usepackage{amsthm}
\usepackage{tikz}

\usepackage{dcolumn}
\usepackage{booktabs}
\usepackage{rotating}
\usepackage{caption,subcaption}
\usepackage{bm}
\usepackage{color}

\usepackage{natbib}

\usepackage{times}
\usepackage{multirow}
\usepackage{setspace}
\usepackage{cases}
\usepackage{txfonts}

\newcommand{\np}[2]{\ensuremath{\langle\,#1\mid #2 \,\rangle}}
\newcommand{\p}[1]{\ensuremath{\langle\, #1\,|}}
\newcommand{\cd}[1]{\ensuremath{|#1\,\rangle}}
\newcommand{\y}[1]{\ensuremath{y^{#1}_{\multimap}}}
\newcommand{\s}[1]{\ensuremath{s^{#1}_{\multimap}}}
\newcommand{\iy}[1]{\ensuremath{y^{\multimap}_{#1}}}


%% file: appendix.tex
\appendix

\section{``Bra-ket" notations for probability formulas}\label{AP1}

This appendix introduces a convenient mathematical representation of probability distribution functions and of conditional density functions using notations involving bra-vector and ket-vector, according to the following definition:
\begin{enumerate}
\item [1a.] A single "bra-vector" $\p{A}$ means the unconditional probability for $A$.
\item [1b.] A single "ket-vector" $\cd{B}$ indicates the  conditions where event $B$ is assumed to happen.
\item [2a.] Scalar-multiplication for ``bra-vector" satisfies the commutative law: $k\cdot\p{A}=\p{A}\cdot k, k\in \mathbb{R}$
\item [2b.] Scalar-multiplication for ``ket-vector" satisfies the commutative law: $k\cdot \cd{B}=\cd{B}\cdot k, k\in \mathbb{R}$ 
\item [3$\phantom{a.}$] The multiplication between a single ``bra" and a single ``ket" vector makes a complete pairwise bra-ket $\np{A}{B}$, which represents the conditional distribution $f(A\mid B)$. The bra-ket  can be regarded as a number.
\item [4a.] The notation $\cd{A}\p{A}$ to concatenate a single ``ket" vector and a single ``bra" vector defines the operator to transform a ``ket" (``bra") into another ``ket" (``bra"), which leads to
\begin{itemize}
\item $\p{B}\,\bigl(\,\cd{A}\p{A}\,\bigr)=\np{B}{A}\p{A}=\p{BA}$
\item $\bigl(\,\cd{A}\p{A}\,\bigr)\,\cd{B}=\cd{AB}\cdot \np{A}{B}$
\end{itemize} 
\item [4b.] Given $A=\bigcup\limits_{i}A_i$, then $\sum_i\,\,\cd{A_i}\p{A_i}=\cd{1}\p{1} := I$ where $I=\cd{1}\p{1}$ denotes 
the identical operator 
such that $\p{B} ( \cd{1}\p{1})= (\np{B}{1}) \p{1} = f(B |1) \p{1}$, where $1$ denotes the whole universe, and thus 
$ \p{1} = p(1)=1$ and $f(B |1) := f(B) = \p{B}$ so that $\p{B} (\cd{1}\p{1})=\p{B}$.
\end{enumerate}

Further, according to the previous defined notations, the classic conditional probability and Bayesian formula can be rewritten as 
\begin{itemize}
\item ({\bfseries Conditional probability formula}): $\np{A}{B}=\dfrac{\p{AB}}{\p{B}}\quad$ or $\quad\p{AB}=\np{A}{B}\p{B}$
\item ({\bfseries The Bayesian formula}): $\qquad\qquad$ $\np{A}{B}=\p{A}\dfrac{\np{B}{A}}{\p{B}}$
\end{itemize}

With the listed formal rules and the basic formula, we can easily give the derivations for the following famous formula in probability theory.
\begin{enumerate}
\item ({\itshape The Law of Total probability\/}): $\p{A}=\p{A}\,I=\p{A}\cdot\sum_i\cd{B_i}\p{B_i}=\sum_i\np{A}{B_i}\p{B_i}$
\item ({\itshape Decomposed Conditional Joint Density\/}): 
\begin{align*}
\np{A,B}{C}&=\p{AB}\cdot\cd{C}=(\np{A}{B}\p{B})\cdot\cd{C}\\
&=\p{A}\,\bigl(\cd{B}\p{B}\,\bigr)\,\cd{C}=\p{A}\cdot\cd{BC}\np{B}{C}\\
&=\np{A}{B,C}\np{B}{C}
\end{align*}
\item ({\itshape Multivariate Bayesian Formula}):
\begin{align*}
\np{A}{B,C}&=\frac{\np{A}{BC}\np{B}{C}}{\np{B}{C}}=\frac{\p{A}\,(\,\cd{B}\p{B}\,)\,\cd{C}}{\np{B}{C}}\\
&=\frac{\p{AB}\cdot\cd{C}}{\np{B}{C}}=\frac{\p{BA}\cdot\cd{C}}{\np{B}{C}}=\frac{\p{B}\,(\,\cd{A}\p{A}\,)\,\cd{C}}{\np{B}{C}}\\
&=\np{A}{C}\frac{\np{B}{A,C}}{\np{B}{C}}
\end{align*}
\end{enumerate}

\section{Proof of eq.\eqref{E:tpfb}}\label{AP2}
\begin{proof}
Since $p_t^{-n}\mid p_{t-1}^{-n}\,\sim\,\mathcal{N}(p_{t-1}^{-n}-n\mu_1,n^2\sigma_1^2)$, one can get
\begin{equation}\label{E:AP21}
\np{p_t^{-n}}{p_{t-1}^{-n}}=\dfrac{1}{\sqrt{2\pi}\sigma_1}\exp\left[-\dfrac12\cdot\left(\dfrac{p_t^{-n}-p_{t-1}^{-n}-\mu_1}{\sigma_1}\right)^2\right]~.
\end{equation}
With the density transformation formula $f_{\tilde{x}|\tilde{y}}(x|y)=f_{\tilde{x}|\tilde{y}}(\phi(x)|y)|\phi'(x)|$, or equivalently \\
$\np{x}{y}=\np{\phi(x)}{y}\cdot|\phi'(x)|$, eq.\eqref{E:AP21} further leads to
\begin{equation}
\np{p_t}{p_{t-1}}=\dfrac{1}{\sqrt{2\pi}\cdot n\sigma_1}\exp\left[-\dfrac12\cdot\left(\dfrac{p_t^{-n}-p_{t-1}^{-n}+n\mu_1}{n\sigma_1}\right)^2\right]\cdot n p_t^{-(n+1)}
\end{equation}
From the definition $y_t=\ln p_t$, and using the density transformation formula again, we obtain
\begin{align}
\np{y_t}{y_{t-1}}&=\np{e^{y_t}}{y_{t-1}}e^{y_t}=\np{p_t}{y_t}p_t\\
&=\dfrac{1}{\sqrt{2\pi}\cdot n\sigma_1}\exp\left[-\dfrac12\cdot\left(\dfrac{e^{-ny_t}-e^{-ny_{t-1}}+n\mu_1}{n\sigma_1}\right)^2\right]\cdot n e^{-ny_t}
\end{align}
Consequently, we have
\begin{equation}
\ln\np{y_t}{1,1,y_{t-1}}=-\dfrac{1}{2}\ln 2\pi-\ln n\sigma_1-\dfrac12\cdot\left(\dfrac{e^{-ny_t}-e^{-ny_{t-1}}+n\mu_1}{n\sigma_1}\right)^2+\ln n -n y_t
\end{equation}
\end{proof}

\section{Derivations for the equations in EM algorithm}
\subsection{proof of the eq. \eqref{E:estep1} from eq.\eqref{E:estep}}\label{AP3.1}
\begin{proof}
\begin{align}
\mathcal{\ln L}_{\pmb{\theta}\,\mid\,\pmb{\theta}^{(k-1)}}&=\sum_{1\cdots T}\ln\p{\y{T},\s{T}}_{\pmb{\theta}}\,\,\p{\y{T},\s{T}}_{\pmb{\theta}^{(k-1)}}=\sum_{1\cdots T}\ln (\,\np{\y{T}}{\s{T}}_{\pmb{\theta}}\p{\s{T}}_{\pmb{\theta}}\,)\,\,\p{\y{T},\s{T}}_{\pmb{\theta}^{(k-1)}}\notag\\
&=\sum_{1\cdots T}\ln \np{\y{T}}{\s{T}}_{\pmb{\theta}}\,\,\p{\y{T},\s{T}}_{\pmb{\theta}^{(k-1)}}+\sum_{1\cdots T}\ln \p{\s{T}}_{\pmb{\theta}}\,\,\p{\y{T},\s{T}}_{\pmb{\theta}^{(k-1)}}\label{E:fisttui}
\end{align}
For the first item in the r.h.s. of \eqref{E:fisttui}, we can further obtain
\begin{align*}
\sum_{1\cdots T}\ln \np{\y{T}}{\s{T}}_{\pmb{\theta}}\,\,\p{\y{T},\s{T}}_{\pmb{\theta}^{(k-1)}}
&=\sum_{1\cdots T}\left(\sum_{t=1}^T\ln \np{y_{t}}{s_t,s_{t-1},\y{t-1}}_{\pmb{\theta}}\right)\,\,\p{\y{T},\s{T}}_{\pmb{\theta}^{(k-1)}}\\
&=\sum_{t=1}^T\left(\sum_{t,t-1}\sum_{/t,/t-1}\ln \np{y_{t}}{s_t,s_{t-1},\y{t-1}}_{\pmb{\theta}}\,\,\p{\y{T},\s{T}}_{\pmb{\theta}^{(k-1)}}\right)\\
&=\sum_{t=1}^T\sum_{t,t-1}\left(\ln \np{y_{t}}{s_t,s_{t-1},\y{t-1}}_{\pmb{\theta}}\,\sum_{/t,/t-1}\p{\y{T},\s{T}}_{\pmb{\theta}^{(k-1)}}\right)\\
&=\sum_{t=1}^T\left(\sum_{t,t-1}\ln \np{y_{t}}{s_t,s_{t-1},\y{t-1}}_{\pmb{\theta}}\,\p{\y{T},s_{t},s_{t-1}}_{\pmb{\theta}^{(k-1)}}\right)\\
&=\p{\y{T}}_{\pmb{\theta}^{(k-1)}} \sum_{t=1}^T\left(\sum_{t,t-1}\ln \np{y_{t}}{s_t,s_{t-1},\y{t-1}}_{\pmb{\theta}}\,\np{s_{t-1},s_t}{\y{T}}_{\pmb{\theta}^{(k-1)}}\right)
\end{align*}
where the notation $\displaystyle \sum_{/t,/t-1}$ represents the summation for all states $s_{\tau}, \tau=1,\cdots,T$,  except for $s_{t-1}$ and $s_t$. The second term in the r.h.s. of \eqref{E:fisttui} leads to
\begin{align*}
\sum_{1\cdots T}\ln \p{\s{T}}_{\pmb{\theta}}\,\,\p{\y{T},\s{T}}_{\pmb{\theta}^{(k-1)}}
&=\sum_{1\cdots T}\left(\sum_{t=1}^T\ln \np{s_t}{s_{t-1}}_{\pmb{\theta}}\right)\,\,\p{\y{T},\s{T}}_{\pmb{\theta}^{(k-1)}}\\
&=\sum_{t=1}^T\left(\sum_{t,t-1}\sum_{/t,/t-1}\ln \np{s_t}{s_{t-1}}_{\pmb{\theta}}\,\,\p{\y{T},\s{T}}_{\pmb{\theta}^{(k-1)}}\right)\\
&=\sum_{t=1}^T\sum_{t,t-1}\left(\ln\np{s_t}{s_{t-1}}_{\pmb{\theta}}\sum_{/t,/t-1}\p{\y{T},\s{T}}_{\pmb{\theta}^{(k-1)}}\right)\\
&=\p{\y{T}}_{\pmb{\theta}^{(k-1)}}\sum_{t=1}^T\left(\sum_{t,t-1}\ln\np{s_t}{s_{t-1}}_{\pmb{\theta}}\,\np{s_{t-1},s_t}{\y{T}}_{\pmb{\theta}^{(k-1)}}\right)
\end{align*}
Thus, we finally obtain \eqref{E:estep1}
\begin{equation*}
\mathcal{\ln L}_{\pmb{\pmb{\theta}}^{(k)}\,\mid\,\pmb{\theta}^{(k-1)}}=
\p{\y{T}}_{\pmb{\theta}^{(k-1)}}\cdot\left(\sum_{t=1}^T \sum_{\substack{\;\;s_t=1,0\\ \!\!s_{t-1}=1,0}}\Bigl(\ln \np{y_{t}}{s_t,s_{t-1},y_{t-1}}_{\pmb{\theta}^{(k)}}+\ln\np{s_t}{s_{t-1}}_{\pmb{\theta}^{(k)}}\,\Bigr)\np{s_{t-1},s_t}{\y{T}}_{\pmb{\theta}^{(k-1)}}\right)
\end{equation*} 
\end{proof}

\subsection{Derivation of the formulas to determine the first four parameters of the HMM}\label{AP3.2}
Since $\p{\s{T}}_{\pmb{\theta}}$ does not depend on $\pmb{\theta}=(\mu_0,\sigma_0, \mu_1, \sigma_1, n)$, one can ignore the second term in eq. \eqref{E:fisttui} when calibrating the parameters of the HMM. 

In order to compute $\mu_0^{(k)}$, the first-order condition of the first term in  eq. \eqref{E:fisttui} reads
\begin{equation}
\sum_{t=1}^T \left(\sum_{\substack{s_t=1,0\\ s_{t-1}=1,0}}\dfrac{\partial}{\partial \mu_0}\ln \np{y_{t}}{s_t,s_{t-1},y_{t-1}}_{\pmb{\theta}}\,\np{s_t,s_{t-1}}{\y{T}}_{\pmb{\theta}^{(k-1)}}\right)=0
\end{equation}
which leads to
\begin{align*}
&\sum_{t=1}^T \left(\dfrac{y_t-y_{t-1}-\mu_0}{\sigma_0}\,\np{s_t=0,s_{t-1}=0}{\y{T}}_{\pmb{\theta}^{(k-1)}}\right)=0\\
&\Longrightarrow\qquad \mu_0=\dfrac{\sum_{t=1}^T (y_t-y_{t-1})\np{s_t=0,s_{t-1}=0}{\y{T}}_{\pmb{\theta}^{(k-1)}}}{\sum_{t=1}^T \np{s_t=0,s_{t-1}=0}{\y{T}}_{\pmb{\theta}^{(k-1)}}}\\
&\Longrightarrow \qquad \mu_0^{(k)}=\dfrac{\sum_{t=1}^T \omega_{(0,0);t}^{(k-1)}(y_t-y_{t-1})}{\sum_{t=1}^T\omega_{(0,0);t}^{(k-1)}}
\end{align*}

In order to compute $\sigma_0^{(k)}$, the first-order condition of the first term in  eq. \eqref{E:fisttui} reads
\begin{equation}
\sum_{t=1}^T \left(\sum_{\substack{s_t=1,0\\ s_{t-1}=1,0}}\dfrac{\partial}{\partial \sigma_0}\ln \np{y_{t}}{s_t,s_{t-1},y_{t-1}}_{\pmb{\theta}}\,\np{s_t,s_{t-1}}{\y{T}}_{\pmb{\theta}^{(k-1)}}\right)=0
\end{equation}
which leads to
\begin{align*}
\sum_{t=1}^T \left(-\dfrac{1}{\sigma_0}+\dfrac{(y_t-y_{t-1}-\mu_0)^2}{\sigma_0^3}\,\right)\np{s_t=0,s_{t-1}=0}{\y{T}}_{\pmb{\theta}^{(k-1)}}=0\\
\Longrightarrow\qquad \sigma_0^{(k)}=\sqrt{\dfrac{\sum_{t=1}^T \omega_{(0,0);t}^{(k-1)}(y_t-y_{t-1}-\mu_0^{(k)})^2}{\sum_{t=1}^T\omega_{(0,0);t}^{(k-1)}}}&
\end{align*}

In order to compute  $\mu_1^{(k)}$, the first-order condition of the first term in  eq. \eqref{E:fisttui} reads
\begin{equation}
\sum_{t=1}^T \left(\sum_{\substack{s_t=1,0\\ s_{t-1}=1,0}}\dfrac{\partial}{\partial \mu_1}\ln \np{y_{t}}{s_t,s_{t-1},y_{t-1}}_{\pmb{\theta}}\,\np{s_t,s_{t-1}}{\y{T}}_{\pmb{\theta}^{(k-1)}}\right)=0
\end{equation}
which leads to
\begin{align*}
& \sum_{t=1}^T \left(-\dfrac{p_t^{-n}-p_{t-1}^{-n}+n \mu_1}{n\sigma_1^2}\,\right)\np{s_t=1,s_{t-1}=1}{\y{T}}_{\pmb{\theta}^{(k-1)}}=0\\
&\Longrightarrow \qquad  \sum_{t=1}^T \left(-p_t^{-n}+p_{t-1}^{-n}-n \mu_1\,\right)\omega_{(1,1);t}^{(k-1)}=0\\
&\Longrightarrow \qquad \mu_1^{(k)}=\dfrac{\sum_{t=1}^T \omega_{(1,1);t}^{(k-1)}(p_{t-1}^{-n}-p_t^{-n})}{n \sum_{t=1}^T\omega_{(1,1);t}^{(k-1)}}
\end{align*}

In order to compute $\sigma_1^{(k)}$, the first-order condition of the first term in  eq. \eqref{E:fisttui} reads 
\begin{equation}
\sum_{t=1}^T \left(\sum_{\substack{s_t=1,0\\ s_{t-1}=1,0}}\dfrac{\partial}{\partial \sigma_1}\ln \np{y_{t}}{s_t,s_{t-1},y_{t-1}}_{\pmb{\theta}}\,\np{s_t,s_{t-1}}{\y{T}}_{\pmb{\theta}^{(k-1)}}\right)=0
\end{equation}
which leads to
\begin{align*}
&\sum_{t=1}^T \left(\dfrac{1}{\sigma_1}-\dfrac{(p_t^{-n}-p_{t-1}^{-n}+n \mu_1)^2}{n^2\sigma_1^3}\,\right)\np{s_t=1,s_{t-1}=1}{\y{T}}_{\pmb{\theta}^{(k-1)}}=0\\
&\Longrightarrow \qquad \sum_{t=1}^T \left[(p_t^{-n}-p_{t-1}^{-n}+n \mu_1)^2-n^2\sigma_1^2\,\right]\omega_{(1,1);t}^{(k-1)}=0\\
&\Longrightarrow  \qquad \sigma_1^{(k)}=\sqrt{\dfrac{\sum_{t=1}^T \omega_{(1,1);t}^{(k-1)}(p_{t}^{-n}-p_{t-1}^{-n}+n\mu_1^{(k)})^2}{n^2 \sum_{t=1}^T\omega_{(1,1);t}^{(k-1)}}}
\end{align*}

\subsection{Derivation for eq.\eqref{E:solven} to solve for $n$}\label{AP3.3}
To compute $n^{(k)}$, one should solve the first-order condition of the first term in  eq. \eqref{E:fisttui} with respect to $n$, that is  
\begin{equation}
\sum_{t=1}^T \left(\sum_{\substack{s_t=1,0\\ s_{t-1}=1,0}}\dfrac{\partial}{\partial n}\ln \np{y_{t}}{s_t,s_{t-1},y_{t-1}}_{\pmb{\theta}}\,\np{s_t,s_{t-1}}{\y{T}}_{\pmb{\theta}^{(k-1)}}\right)=0
\end{equation}
which gives, 
\begin{multline}
 \sum_{t=1}^T \Biggl(
-\dfrac{(p_t^{-n}-p_{t-1}^{-n}+n\mu_1)(-p_t^{-n}\ln p_t+p_{t-1}^{-n}\ln p_{t-1}+\mu_1)}{n^2\sigma_1^2}\\+\dfrac{(p_t^{-n}-p_{t-1}^{-n}+n\mu_1)^2}{n^3\sigma_1^2}-\frac{1}{n}+\frac{1}{n}-\ln p_t\Biggr)\,\,\omega_{(1,1):t}^{(k-1)}=0
\end{multline}

Note that $\displaystyle \sum_{t=1}^T \left(\dfrac{(p_t^{-n}-p_{t-1}^{-n}+n\mu_1)^2}{n^3\sigma_1^2}-\frac{1}{n}\right)\,\omega_{(1,1):t}^{(k-1)}=0$. Thus, the above equation can be reduced to
\begin{equation}
\sum_{t=1}^T \left(-\dfrac{(p_t^{-n}-p_{t-1}^{-n}+n\mu_1)(-p_t^{-n}\ln p_t+p_{t-1}^{-n}\ln p_{t-1}+\mu_1)}{n^2\sigma_1^2}+\frac{1}{n}-\ln p_t\right)\,\omega_{(1,1):t}^{(k-1)}=0\label{2}
\end{equation}

\subsection{Derivation of the formulas to compute the optimal posterior transition probability from $s_{t-1}$ to $s_t$}\label{AP3.4}

Since $\np{s_t}{s_{t-1}}_{\pmb{\theta}^{(k)}}$ is only related to the second term in eq.\eqref{E:fisttui}, we can ignore the first term when performing the optimization:
\begin{align*}
\sum_{1\cdots T}\ln \p{\s{T}}_{\pmb{\theta}}\,\,\p{\y{T},\s{T}}_{\pmb{\theta}^{(k-1)}}&=
\sum_{t=1}^T\left(\sum_{t,t-1}\ln\np{s_t}{s_{t-1}}_{\pmb{\theta}}\,\np{s_{t},s_{t-1}}{\y{T}}_{\pmb{\theta}^{(k-1)}}\right)\\
&=\sum_{t=1}^T\left(\sum_{j,i}\ln q_{ij}^{(k)}\,\np{j,i}{\y{T}}_{\pmb{\theta}^{(k-1)}}\right)
\end{align*} 
Without loss of generality, the first-order condition against $q_{i1}$ leads to
\begin{align*}
0&=\sum_{t=1}^T\left(\dfrac{\partial}{\partial q_{i1}^{(k)} }\ln q_{ij}^{(k)}\,\np{i,j}{\y{T}}_{\pmb{\theta}^{(k-1)}}\right) \\
&=\sum_{t=1}^T\left(\dfrac{1}{q_{i1}^{(k)}}\np{1,i}{\y{T}}_{\pmb{\theta}^{(k-1)}}-\dfrac{1}{1-q_{i1}^{(k)}}\np{0,i}{\y{T}}_{\pmb{\theta}^{(k-1)}}\right)
\end{align*}
Then we have 
\begin{align*}
&q_{i1}^{(k)}=\dfrac{\sum_{t=1}^T\np{1,i}{\y{T}}_{\pmb{\theta}^{(k-1)}}}{\sum_{t=1}^T(\np{1,i}{\y{T}}_{\pmb{\theta}^{(k-1)}}+\np{0,i}{\y{T}}_{\pmb{\theta}^{(k-1)}})}=\dfrac{\sum_{t=1}^T\np{1,i}{\y{T}}_{\pmb{\theta}^{(k-1)}}}{\sum_{t=1}^T\np{i}{\y{T}}_{\pmb{\theta}^{(k-1)}}}
\end{align*}
It is the optimal posterior transition probability after the $k$th iteration, $q_{ij}^{(k)}$, when $j=1$.

\subsection{Reasoning to replace $\np{s_t}{s_{t+1},\y{t},\iy{t+1}}$ by $\np{s_t}{s_{t+1},\y{t}}$ in eq.\eqref{E:dddlla}} \label{AP3.5}
\begin{align}
\np{s_t}{s_{t+1},\y{t},\iy{t+1}}&=\frac{\p{s_t}\,\biggl(\,\cd{s_{t+1},\y{t},\iy{t+1}}\,\biggr)\,\np{\iy{t+1}}{s_{t+1},\y{t}}}
{\np{\iy{t+1}}{s_{t+1},\y{t}}}\\
&=\frac{\p{s_t}\,\biggl(\,\cd{s_{t+1},\y{t},\iy{t+1}}\,\np{\iy{t+1}}{s_{t+1},\y{t}}\,\biggr)}
{\np{\iy{t+1}}{s_{t+1},\y{t}}}\\
&=\frac{\p{s_t}\,\biggl(\,\cd{\iy{t+1}}\p{\iy{t+1}}\,\biggr)\,\cd{s_{t+1},\y{t}}}
{\np{\iy{t+1}}{s_{t+1},\y{t}}}\\
&=\frac{\biggl(\,\np{s_t}{\iy{t+1}}\p{\iy{t+1}}\,\biggr)\,\cd{s_{t+1},\y{t}}}{\np{\iy{t+1}}{s_{t+1},\y{t}}}=\frac{\p{\iy{t+1}}\,\biggl(\,\cd{s_t}\p{s_t}\,\biggr)\,\cd{s_{t+1},\y{t}}}{\np{\iy{t+1}}{s_{t+1},\y{t}}}\\
&=\frac{\np{\iy{t+1}}{s_t,s_{t+1},\y{t}}\np{s_t}{s_{t+1},\y{t}}}{\np{\iy{t+1}}{s_{t+1},\y{t}}}\label{E:opa51}
\end{align}

By assuming that the joint distribution of $( y_{t+1},y_{t+2},\cdots,y_{T} )$ is irrelevent to $s_t$, we can eliminate the denominator and easily recover \eqref{E:opa51} for $\np{s_t}{s_{t+1},\y{t}}$